\documentclass[a4paper]{article}
\usepackage[english]{babel}
\usepackage[utf8x]{inputenc}
\usepackage[T1]{fontenc}
\usepackage{authblk}
\usepackage{adjustbox}

\usepackage{booktabs}       %
\usepackage{amsfonts}       %
\usepackage{nicefrac}       %
\usepackage{microtype}      %
\usepackage{diagbox} %
\usepackage{dsfont}
\usepackage{bm} %

\usepackage{amsmath}
\usepackage{graphicx}
\usepackage{amsfonts}
\usepackage{subcaption}
\usepackage{stmaryrd}
\usepackage{color, colortbl}
\usepackage{breakcites}
\usepackage{pdfpages}
\usepackage{rotating}
\usepackage{dirtree} %
\usepackage{multirow}
\usepackage{soul}
\usepackage{color, colortbl} %
\usepackage{multicol}
\usepackage{amssymb}
\usepackage{enumitem}
\definecolor{Gray}{gray}{0.9}
\definecolor{LightCyan}{rgb}{0.88,1,1}  %

\usepackage{array}

\newcolumntype{P}[1]{>{\centering\arraybackslash}p{#1}}

\usepackage[colorlinks]{hyperref}

\usepackage{hyperref}
\hypersetup{
     colorlinks   = true,
     citecolor    = blue
}
\usepackage{pdflscape}

 \usepackage[colorinlistoftodos]{todonotes}

\usepackage{makecell}
\usepackage{rotating}
\usepackage{adjustbox}

\newtheorem{mydef}{Definition}
\newtheorem{rec}{Recommendation}
\usepackage{graphicx}
\usepackage{lscape}
\usepackage{wasysym}
\usepackage{ textcomp } 
\usepackage{xurl}

\title{\textit{Questioning causality on sex, gender and COVID-19, and 
identifying bias in large-scale data-driven analyses: 
the Bias Priority Recommendations} and \textit{Bias Catalog for Pandemics}}

\date{\vspace{-5ex}}

\author[1]{Natalia D\'iaz-Rodr\'iguez}
\author[2]{R\=uta Binkyt\.e-Sadauskien\.e}
\author[3]{Wafae Bakkali}
\author[4]{Sannidhi Bookseller}
\author[5]{Paola Tubaro}
\author[6]{Andrius Bacevi\v{c}ius}
\author[7]{Raja Chatila}
\affil[1]{
U2IS, Inria Flowers team, ENSTA Paris, Institut Polytechnique Paris, 
Palaiseau, France.}%
\affil[2]{Inria Saclay-Ile-de-France Comete team, LIX, 
Palaiseau, France.}%
\affil[3]{DataLab Group, Cr\'edit Agricole, 
Montrouge, France.}%
\affil[4]{
EPITA College, 
Le Kremlin-
Bic\^etre, France.}%
\affil[5]{LISN-TAU, CNRS, University Paris-Saclay, Inria, France.}
\affil[6]{OSE Immunotherapeutics, Paris, France.}
\affil[7]{ISIR (Institute of Intelligent Systems and Robotics), Sorbonne University, Paris, France.}

\providecommand{\Keywords}[1]{\textbf{\textit{Keywords: }} #1}
\begin{document}
\maketitle

\begin{abstract}
The COVID-19 pandemic has spurred a large amount of observational studies reporting 
linkages between the risk of developing severe COVID-19 or dying from it, and sex and gender.
By reviewing a large body of related literature and conducting a 
fine grained analysis based on sex-disaggregated data of 61 countries spanning 5 continents, we discover several confounding 
factors that could possibly explain the supposed male vulnerability to COVID-19. We thus highlight the challenge of making causal claims based on available data, given the lack of statistical significance and potential existence of 
biases. 
Informed by our findings on potential variables acting as confounders, we contribute a broad overview on the issues bias, explainability and fairness entail in data-driven analyses. 
Thus, we outline a set of discriminatory policy consequences that could, based on such results, lead to unintended discrimination.
To raise awareness on the dimensionality of such foreseen impacts, we have compiled an encyclopedia-like reference guide, the \textit{Bias Catalog for Pandemics} (BCP), to provide definitions and emphasize realistic examples of bias in general, and within the COVID-19 pandemic context. 
These are categorized within a division of bias families and a 2-level priority scale, together with preventive steps. 
In addition, we facilitate the \textit{Bias Priority Recommendations} on how to best use and apply this catalog, and provide guidelines in order to address real world research questions. The objective is to anticipate and avoid disparate impact and discrimination, by considering causality, explainability, bias and techniques to mitigate the latter. 
With these, we hope to 1) contribute to designing and conducting fair and equitable data-driven studies and research; and 2) interpret and draw meaningful and actionable conclusions from these. 


\end{abstract}





\section{Introduction}
Sex and gender disparity was noticed in many cases of Coronavirus disease 2019 (COVID-19). In this article we follow the definition proposed by \cite{ahmed2020sex} distinguishing between \textit{sex} as a set of biological attributes, and \textit{gender} as a social-psychological category. As it is later demonstrated, both might have an impact on COVID-19 mortality rates. We also note that in this study we consider binary values for gender, although an in depth analysis of gender roles would potentially yield a more complex picture. The disease 
is reported to be deadlier for infected men than women with a 2.8\% fatality rate in Chinese men versus 1.7\% in women \cite{gebhard2020impact}, while sex-disaggregated data for COVID-19 in several European countries shows a similar number of cases between sexes, but more severe outcomes in aged men \cite{gebhard2020impact}.


Biological differences in the immune system in men and women may affect the person's ability to fight COVID-19.
It may be argued that men are more vulnerable to the COVID-19 in relation to women because of a distinctive lifestyle, smoking, drinking, working hours, sex hormones, hypertension, and other circumstances. Physicians have struggled to understand the disease and to come up with a unified theory for how it works \cite{smith2020supercomputer}. 
Data regarding sex differences in ACE2 and TMPRSS2 enzymes are not yet coherent 
and the link between circulating ACE2 and COVID-19 is not clear \cite{gebhard2020impact}. 
Additionally, more research is required on how sex and gender intersects with age and race, to further increase the risk of severe COVID-19 outcomes in men. 

In \textit{PLoS pathogens} and \textit{CMAJ} journals 
it is also discussed how other socio-economic factors also increase the risk of COVID-19 \cite{klein2020biological, tadiri2020influence}. 
Systemic health and social inequities have disproportionately exposed low-income communities, racial and ethnic minorities to higher risk of COVID-19 infection and death. Additionally, 
uneven testing strategies 
across the world, and the quality of epidemiological data, limit the accuracy of estimated distribution of COVID-19 patients according to \cite{kopel2020racial}.


These studies call for an information bias check. Information bias (or misinformation) arises from systematic differences in the collection, recall, recording or handling of information used in a study \cite{catalogOfBias17}. It is one of the most common sources of bias compromising the validity of health research and it emerges from the approaches used to obtain or confirm study measurements. These measurements can be obtained by experimentation (such as bioassays) or observation (e.g., questionnaires) \cite{althubaiti2016information}.

Bias can be defined as any systematic error in the design, conduct, or analysis of a study. In health studies, bias can arise from two different sources; the approach adopted for selecting subjects for a study or the approach adopted for collecting or measuring data from a study. These are, respectively, termed as selection bias and information bias. 
Bias can have different effects on the validity of medical research findings. In epidemiological studies, bias can lead to inaccurate estimates of association, or over- or underestimation of risk parameters. 
Allocating time and resources to consider bias and their impacts on final results is key for avoiding pitfalls and making valid conclusions \cite{althubaiti2016information}.

In this paper we focus on analyzing bias in clinical studies and data-driven, i.e., observational studies, and not necessarily bias in machine learning and artificial intelligence. However, this is a preliminary step that can further propagate more intricate forms of bias in computational models to deal with pandemic policies and decision making. By \textit{data-driven} decisions we will refer both to predictions learned from data and policies or clinician's decisions based on those outcomes. With respect to the clinical healthcare domain, the same notion will be termed as \textit{observational} studies, and thus, in this paper we consider both data-driven and observational as synonyms. 

As a result, we bring light into: 1) a potential set of hypotheses within our COVID-19 case study to further verify its causal link, and 2) the unintended consequences that can derive from a lack of an adequate toolbox to mitigate bias. 

The contributions of this paper are the following:
\begin{itemize}
    \item A review of the most up-to-date  
    literature mainly from 2020-2021 analysing observational studies, from the gender and sex perspective, that indicate an increased male vulnerability to COVID-19.  
    \item An identification of a set of hypotheses that can potentially be responsible for the observed disparities.
    \item The former factors are further investigated through illustrated factor graphs detailing the unintended discrimination policies they can lead to.  
    \item 
    We highlight explainability and causality instrumental approaches to better understand the data and facilitate equitable data-driven decisions.
    \item A synthesis of our findings is contributed in the form of the \textit{Bias Catalog for Pandemics} and \textit{Bias Priority Recommendations}, which review the minimal ingredients necessary to account for when mitigating potential sources of bias possibly explaining reported disparities. 
      
\end{itemize}


The rest of this paper is organized as follows. First we present the most recent literature on gendered and sex-related effects of COVID-19 in Section \ref{sec:relatedWork}. In Section \ref{sec: diggingDeeper} we conducted a data analysis on publicly available data to investigate the possible impact of 
sex-related confounding factors on the COVID-19 outcomes. We discern several factors where the reports could lead to discrimination in Section \ref{sec:fair}. Section \ref{sec:bias_priority} presents prioritized check-lists for detecting and handling bias. 
Finally, we discuss results and open research directions for the future in Sections \ref{sec:discussion} and \ref{sec:conclusions} respectively. 

 \section{Related Work: 
 Identifying bias on large-scale 
 data-driven studies 
 on COVID-19 }
 \label{sec:relatedWork} 
 
What describes the data-driven paradigm is that knowledge is obtained by purely statistical analysis of the data, without relying on previous hypotheses \cite{kitchin2014big}. As an example in our case, the association between gender and COVID-19 mortality was not something scientists were looking for based on previous knowledge and hypothesis, but a data-driven discovery. 

 We review data-driven findings on gender and COVID-19 from two angles. First, we analyze a body of papers placing gender and sex as a risk factor towards COVID-19, focusing on explaining the reasons behind disparity.
 Second, we discuss about possible fairness implications of the former results that could lead to discrimination decisions, with the aim of guiding the design of an ultimate overall bias control protocol to be put in place to uncover the potential hidden faults that can carry on too far, leading to 
 inequity.  
 

The amount of literature providing large-scale claims 
is undoubtedly blooming \cite{besserve2021assaying}. The disparity in sex and gender is very important to get a better understanding of the potential impact of gender on COVID-19. 
Vaccines are the best prophylactic treatment for infectious diseases as they provide immunity and protection prior to infection. Sex and gender impact vaccine acceptance, responses, and outcomes \cite{gebhard2020impact}. Women are often less likely to accept vaccines but once vaccinated, develop higher antibody responses \cite{klein2010xs}. For example, after vaccination against influenza, yellow fever, rubella, mumps, measles, small pox, hepatitis A and B and dengue viruses, protective antibody responses are twice as high in adult females compared with males. However they report more adverse reactions to vaccines than males  \cite{gebhard2020impact}.
Moreover, biological differences in the immune systems of men and women exist, and they may affect the capacity to fight COVID-19 infection. The relevance of gender norms, roles, and relations that influence women and men differential vulnerability to infection, exposure to pathogens, treatment received, as well as how these may differ among different groups of women and men is outlined in \cite{wenham2020covid}. 


Men appear to be at a greater risk with COVID-19 compared to women, whose higher immunologic response is probably associated with decreased mortality.
Furthermore, certain differences in cardiac manifestations in COVID-19 must be considered as a core component \cite{sharma2020sex}. 
From the observational studies perspective, men appear to be at a greater risk. 
Table \ref{tab:covidPapers} shows articles finding men to be more vulnerable to COVID-19 in comparison to women. 
Sex is surely not the only risk factor in a disease that, according to \cite{smith2020supercomputer}, is challenging to diagnose and theorize, and whose effects also depend on vulnerabilities related to diabetes, obesity, hypertension, heart disease, chronic kidney disease, and chronic pulmonary disease according to \cite{klein2020biological}.

When comparing the COVID-19 case fatality rate (CFR) 
between China and Italy, the authors in \cite{von2020simpson} infer how methods from causal inference --in particular, mediation analysis--, can be used to resolve apparent statistical paradoxes and other various causal questions from data regarding the current pandemic. Many research studies \cite{head2020effect} revealed that systemic health and social inequities have disproportionately increased the risk of COVID-19 infection and death among low-income communities and racial and ethnic minorities. The outcomes in \cite{bertsimas2020predictions} provide insights on the clinical aspects of the disease, on patients’ infection and mortality risks, on the dynamics of the pandemic, and on the levels that policymakers and healthcare providers can use to alleviate its toll. Many authors suggest that women naturally produce more types of interferon, which limits the abnormal immune response in the form of serious cases of COVID-19. Moreover, women also produce more T lymphocytes which kill infected cells; and the "female" hormone estradiol would also offer greater protection against infection. On the contrary, studies indicate testosterone would limit the immune response in men, which may explain the observed sex-bias \cite{peckham2020male}. 

In the gender and social norms side, a recent study conducted
in Spain (one of the hardest hit countries in Europe) reported that women had more responsible attitude towards the COVID-19 pandemic than men \cite{de2020could}, and another in the US showed that women take more precautions, wear more masks and cover more coughs than men\footnote{\url{https://hbswk.hbs.edu/item/the-covid-gender-gap-why-fewer-women-are-dying}  
} \cite{haischer2020wearing}.


On the other side of the coin, other studies look deeper, paying attention to more fine grained variables. The branch of studies that go deeper on trying to identify further reasons behind these differences on the vulnerability of men and women to critical conditions of COVID-19 are briefly summarized next.

Immunitary response duration was studied at the Pasteur Institut\footnote{\url{https://www.pasteur.fr/fr/espace-presse/documents-presse/COVID-19-duree-reponse-immunitaire-neutralisante-plus-longue-femmes-que-hommes}} and CHU of Strasbourg on 308 healthcare personnel that developed a light form of COVID-19 \cite{grzelak2020sex}. They show significantly steeper, i.e., faster decline in antibodies (anti-S and NAbs) in males than in females independently of age and BMI, hinting to a lower duration of protection after SARS-CoV-2 infection or vaccination. 
As more protective antibodies are formed in women, they last longer and so, women are better protected. 

Despite the virulence of the virus (strength, ability to transmit and infect) being in no way gender dependent, 
two risk factors of older age and concomitant illness, such as those that suppress the immune system, were identified. In such concomitant (heart and metabolic) diseases, gender was not a risk factor and no significant difference was observed in the incidence of complications between men and women \cite{o2021coronavirus}\footnote{\url{https://www.cdc.gov/coronavirus/2019-ncov/need-extra-precautions/index.html?CDC_AA_refVal=https\%3A\%2F\%2Fwww.cdc.gov\%2Fcoronavirus\%}}. 
For instance, a comparative risk assessment analysis framework including probabilistic sensitivity analysis and Monte Carlo simulations to integrate in the data stratum-specific uncertainties, found the major US cardio-metabolic conditions in decreasing order of risk-attribution to be diabetes mellitus, obesity, hypertension, and heart failure \cite{o2021coronavirus}.

%

The hypothesis that testosterone levels can influence clinical worsening and mortality of COVID-19 was analyzed through the study of several biochemical risk factors in a cohort of male patients admitted to respiratory intensive care unit \cite{testoster_study}. Their results show a correlation between lower testosterone levels and transfer to higher care units or death. The study includes control for age, smoking, obesity and health status. Interestingly, the testosterone levels did not vary significantly with age, leading authors to conclude that observed hormonal changes can be a consequence and not a reason for the patients' condition. Therefore it is not clear if the men with lower initial testosterone levels are more vulnerable, or hormonal change is a symptom of worsening due to other conditions. The authors acknowledge that randomized trials with testosterone treatment would be needed to establish a causal direction. The study does not compare the clinical outcomes in men and women directly, but contributes to better understanding of differential impact of COVID-19 on men and women observed in the epidemiological data. 

Another study investigated the potential critical role of testosterone in a higher severe status rate and mortality rate among men tested positive for COVID-19 \cite{testoster2}. The authors showed that men with lower levels of serum T, which is generally associated with ageing, obesity and other chronic diseases, are more prone to develop pulmonary and systemic inflammation and worse respiratory and general parameters. However, results highlighted the fact that more investigations are needed to analyze this biological phenomenon as no clear causal link has been found.

The higher antibody concentration in female may play an important role in preventing patients from progressing into a severe status and even death. While the generation of SARS-CoV-2 IgG antibodies in female patients was stronger than in male patients in the disease early phase, and also in the severe status stage, the concentration of IgG antibody in mild, general, and recovering patients showed (via Mann-Whitney U test) no statistical difference between male and female patients \cite{zeng2020comparison} in a study including 127 male and 204 female hospital patients.


\begin{landscape}
 \begin{table}[htbp!]
  \centering
  \begin{tabular}{ | m{5.5 cm} | m{5.5 cm} | m{2 cm} | m{3 cm} | m{2 cm} | m{2 cm} | }
 
  \hline 
  \textbf{Study} & \textbf{Tested Hypothesis} & \textbf{Men are more vulnerable 
  } 
  & \textbf{Reported Health Conditions} & \textbf{Age Correlation} & \textbf{Reported Drinking / Smoking} 
  \\\hline
  
  \emph{Impact of sex and gender on COVID-19 outcomes in Europe \cite{gebhard2020impact}}
  & COVID-19 is deadlier for infected men than women 
  & \checked (NSD: M) 
  & C, H, DM, CD, CRD, CLD
  & \checked
  & \checked (D, S)
  \\\hline
  
  \emph{Coronavirus: why men are more vulnerable to COVID-19 than women? \cite{bwire2020coronavirus}} 
  & There are higher morbidity and mortality rates in males than females
  & \checked (NSD: M)
  & O, DM, H
  & \checked
  &    \\\hline

  \emph{Biological sex impacts COVID-19 outcomes \cite{klein2020biological}}
  & Mechanistic differences including the expression and activity of ACE2 enzyme 
  result in antiviral immunity, cases, hospitalizations and deaths differences. 
  & \checked (NSD: M)
  & CPD, CKD, II, HD, O
  & \checked
  & 
  \\\hline
  
 \emph{COVID-19: the gendered impacts of the outbreaks \cite{wenham2020covid}}
  & 
  Men are more likely to remain hospitalized an die and less likely to be discharged from the hostpital than women.
  & \checked (NSD: M) 
  & H
  & \checked
  & \checked (S)
  \\\hline
  
  \emph{Racial and gender based differences in COVID-19 \cite{kopel2020racial}}
  & 
  Ethnic differences 
  influence susceptibility and mortality 
  %
  & \checked (NSD: M)
  & HD, O, CLD, C,  H,  DM, CD
  & \checked
  & \checked (D, S)\\
  \hline

  \emph{Sex Differences in Mortality From COVID-19 Pandemic: Are Men Vulnerable and Women Protected? \cite{sharma2020sex}}
  & Male sex plays a role in increased mortality rates 
  & \checked (NSD: M)
  & H,  DM, CD, CRD, CLD
  & \checked
  & 
  \\\hline
  
  \emph{The influence of sex and gender domains on COVID-19 cases and mortality \cite{tadiri2020influence}} & Gender Inequality Index is positively associated with male:female cases ratio 
  & \checked (SSD: M) 
  & 
19

  & 
  &   \\\hline

  \emph{Male sex identified by global COVID-19 meta-analysis as a risk factor for death and ITU admission \cite{peckham2020male}} & Male sex is a risk factor for death and ITU admission but not for infections. 
  & \checked (SSD: M)
  & H, II, C
  & \checked
  & \checked (S)
  \\\hline
 \end{tabular}

 \caption{Summary of 
 claims involving statements regarding men being more affected by the COVID-19 compared to women. X indicates correlation of that variable with the COVID-19. M: men are more affected, F: women are more affected by COVID-19, SSD: Statistically Significant Difference, NSD: Non Statistically-significant difference. Factors: S: Smoking, D: Drinking,  C: Cancer, H: Hypertension, DM: Diabetes mellitus, CD: Cardiovascular diseases, CRD: Chronic respiratory disease, CLD-chronic lung disease, HD: Heart disease, O: Obesity, II: Inflammatory immune responses, CHK: Chronic kidney disease, CPD: Chronic pulmonary disease. 
 Even though most articles claim men are more affected by COVID-19 than women and die more, none of them shows statistical significance nor has enough data to provide causal links beyond correlational studies.}
 \label{tab:covidPapers}

 
 \end{table}
\end{landscape}


Against most of the literature observed, the sex bias observed in COVID-19 as stated by \cite{peckham2020male}, is a worldwide phenomenon suggested by \textit{anecdotal evidence}. 
While there is no difference in the proportion of males and females with confirmed COVID-19, their conclusion does not extend to the severe cases side of the meta-analysis, where male patients have almost three times the odds of requiring intensive treatment unit (ITU) admission \cite{peckham2020male}.


\subsection{
Potential confounders as explanatory variables of sex and gender dependent vulnerability}

From observational analyses and tables in previous section we can observe a set of factors repeating 
as conditioning factors to explain the differences of sex and gender's impact on COVID-19. In this section we synthesize these factors to provide an overall aggregation of COVID-19-related claims most stated by the literature on the impact of different variables on COVID-19.



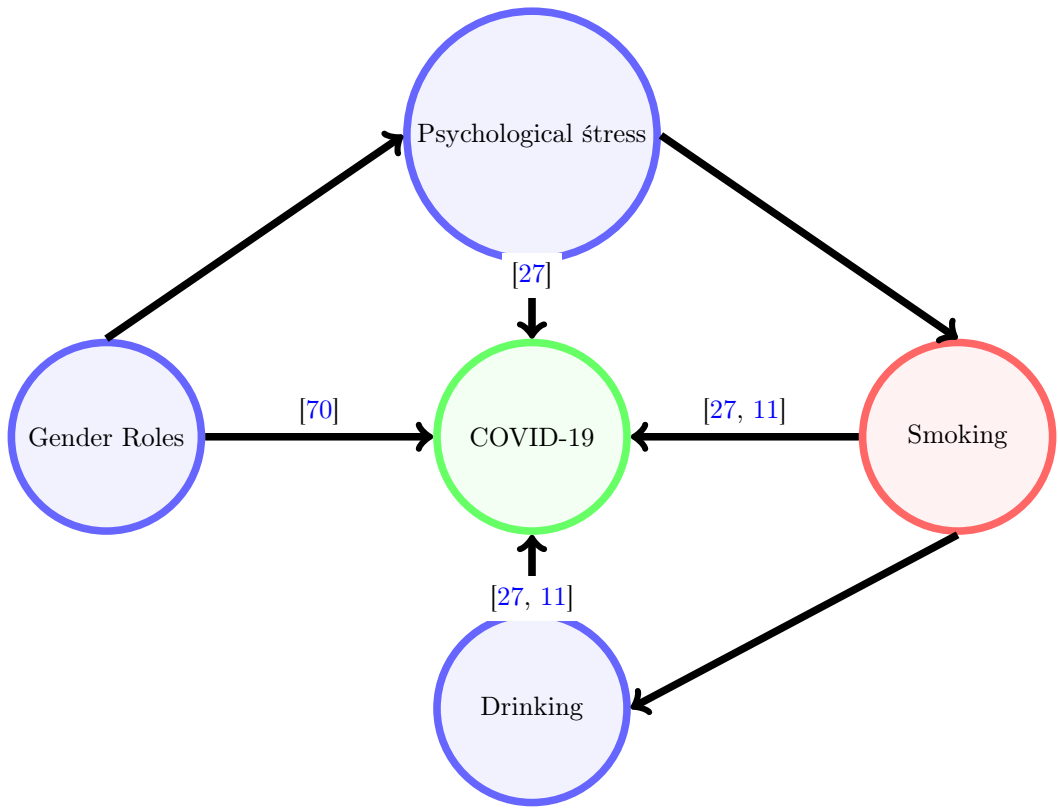
\begin{figure}[htbp!]
\begin{tikzpicture}
[
roundnode/.style={circle, draw=black!60, fill=white!5}
]
\node[roundnode, draw=green!60, fill=green!5,line width=1mm, minimum size= 2.5cm] (maintopic)                        {COVID-19};
\node[roundnode,draw=blue!60, fill=blue!5, line width=1mm, minimum size=2.5cm] (uppercicle) [above=of maintopic]  {Psychological \' stress};
\node[roundnode, draw=red!60, fill=red!5, line width=1mm, minimum size=2.5cm] (rightcicle) [right=3 cm of maintopic]  {Smoking};
\node[roundnode, draw=blue!60, fill=blue!5, line width=1mm, minimum size=2.5cm] (leftcircle) [left= 3 cm of maintopic]   {Gender Roles};
\node[roundnode, draw=blue!60, fill=blue!5, line width=1mm, minimum size=2.5cm] (lowercircle) [below=of maintopic] {Drinking};

\draw[->,line width=1mm] (uppercicle.south) -- (maintopic.north)node [midway,above, fill=white]{\cite{gebhard2020impact}};
\draw[->, line width=1mm] (rightcicle.west) -- (maintopic.east)node [midway,above, fill=white]{\cite{gebhard2020impact, bwire2020coronavirus}};
\draw[->, line width=1mm] (leftcircle.east) -- (maintopic.west)node [midway,above, fill=white]{\cite{tadiri2020influence}};
\draw[->, line width=1mm] (lowercircle.north) -- (maintopic.south)node [midway,below, fill=white]{\cite{gebhard2020impact, 
bwire2020coronavirus}};

\draw[->, line width=1mm] (uppercicle.east) -- (rightcicle.north);
\draw[->, line width=1mm] (rightcicle.south) -- (lowercircle.east);
\draw[->, line width=1mm] (leftcircle.north) -- (uppercicle.west);

\end{tikzpicture}
\caption{Factor graph with most identified correlational variables and potential causes: \textit{Psychological stress, Gender roles} (norms, responsibilities, gender disparity, etc.), and \textit{Drinking} are major factors of COVID-19 infection. Gender differences in risk behaviors, such as smoking and drinking, may be contributing to the gender gaps in mortality of such non-communicable diseases. Factors in \color{red}red \color{black} show the \color{red}most adverse effect \color{black} on COVID-19 impact, while those in \color{blue}blue \color{black} show the \color{blue}affected factors \color{black} to a lesser extent. Being smoker seems to be the most affected factor. All factors are interconnected as well, as they directly lead to COVID-19.}
\label{fig:FactorGraph}
\end{figure}

In form of a factor 
graph in Figure \ref{fig:FactorGraph} we show the role that sex, gender norms and lifestyle (drinking, smoking) can play on getting affected by COVID-19. In red are shown the factors that create some of the hazardous effects of COVID-19 outcomes. The factors in blue are affecting COVID-19 outcomes to some extent. 

Smoking and drinking rates are higher among men than women worldwide, 
which induce lung injuries that affect COVID-19 vulnerability \cite{gebhard2020impact}. 

Gender roles are considered as those influencing women's and men's different vulnerability to infection and exposure to pathogens, as reported in \cite{wenham2020covid}. The impact of gender-specific lifestyle, health behavior, psychological stress, and socioeconomic conditions on COVID-19 is further studied in \cite{gebhard2020impact}.

We are aware that other studies have considered other factors as important ones in the way COVID-19 infection translates into a severe case, for instance, the blood group type \cite{zietz2020associations, zietz2020testing,POURALI2020100743}, vitamin D deficit \cite{ebadi2020perspective,jain2020analysis}, or other genetic factors \cite{zeberg2020major}. However, for the purpose of simplifying our main point of illustrating potential discriminatory policies, in the rest of the paper we will restrict ourselves to highlight the subset of factors, among the many possible ones, summarized in Figure \ref{fig:FactorGraph}. The purpose is illustrating potential discrimination scenarios that could emerge from such suppositional explanations. The factor graph is therefore constructed based on the reported influence directions of those variables found in the reviewed literature. We acknowledge that in reality the formal causality and causability evaluation methods, as described by \cite{Pearl_review} and \cite{holzinger_2020}, respectively, should be applied on the data and the model to ensure proper understanding of the relationship between sex, gender and COVID-19 and enhance the following of fair decision policies. 


Next sections will elaborate on the formal tools available to further study and corroborate such causal hypotheses and explanatory factors drawing on the body of analyzed literature.

\section{
Digging deeper: towards a more accurate assessment of the impact of sex and gender on COVID-19} 

\label{sec: diggingDeeper}

To help understanding the association between gender and COVID-19, we conducted a more focused data analysis based on publicly available data to investigate the possible impact of sex and gender on the COVID-19 epidemic\footnote{Data analysis notebook in R available for reproducibility online: \url{https://rpubs.com/wafaeB/684506}}. We constructed a database that aggregates confirmed cases statistics, COVID-19 deaths, ICU admissions and smoking data per gender for 61 countries spanning 5 continents. Our data sources are briefly described below.
\begin{itemize}
    \item The \textit{Global Health 50/50}\footnote{Global Health 50/50 project website \url{https://globalhealth5050.org/}} project housed at University College of London, which is created by a live tracker that aggregates data on COVID-19 cases and mortality from published government reports. At the time of our analysis on January 05, 2021, sex-disaggregated data for 183 countries including confirmed cases, confirmed deaths, etc. was represented in the live tracker. 
    \item We also used a public dataset maintained by \textit{Our World in Data}\footnote{Our World in Data portal \url{ourworldindata.org}}, which also contains additional information such as smoking, population, and daily COVID-19 cases.
\end{itemize}
By aggregating data from these two sources, and including only countries for which confirmed cases, deaths and smoking information is available, we were able to analyze complete data of 61 countries\footnote{The total aggregated multisource data contained the following countries: Albania, Denmark, France, Germany, Italy, Jamaica, Kyrgystan, Latvia, Moldova, Netherlands, Portugal, Romania, South Africa, South Korea, Spain, Sweden, Switzerland, Tunisia, Ukraine, Australia, Belgium, Canada, Estonia, Finland, Slovenia, Argentina, Austria, Bangladesh, Bosnia and Herzegovina, Burkina Faso, Chile, China, Colombia, Costa Rica, Czech Republic, Dominican Republic, Greece, Ecuador, Haiti, India, Indonesia, Iran, Israel, Kenya, Liberia, Luxembourg, Malawi, Maldives, Mexico, Morocco, Myanmar, Nepal, Norway, Pakistan, Panama, Paraguay, Philippines, Thailand, Turkey, Uganda, Vietnam.}.

We then looked at the \textit{male-to-female} (male/female) ratio of confirmed cases, $\rho_{cases}$, and compared it to the \textit{male-to-female} ratio of deaths, $\rho_{deaths}$, for each country. We particularly classified countries on 4 groups based on these two parameters as follows: 
\begin{itemize}
    \item Group 1 includes the countries in which $\rho_{cases}$ \textgreater $1$ and $\rho_{deaths} \textless 1$. In our analysis, only one country belongs to Group 1.
    \item Group 2, which contains 19 countries, represents countries in which $\rho_{cases} \textless 1$ and $\rho_{deaths} \textgreater 1$.
    \item Group 3 contains 6 countries in which $\rho_{cases} \textless 1$ and $\rho_{deaths} \textless 1$.
    \item Group 4 includes 35 countries in which $\rho_{cases} \textgreater 1$ and $\rho_{deaths}$ \textgreater$ 1$.
\end{itemize}

\begin{table}[htbp!]
\centering
\begin{tabular}{l|l|l}
\hline
\textit{More Deaths: } & Females & Males\\
\textit{More Cases:} & & \\\hline\hline
Females & Group 3 & Group 2 \\
Males  & Group 1 & Group 4 \\ \hline 
\end{tabular}
\caption{\label{tab:groups}Summary of analyzed male-to-female cases ratio and male-to-female deaths ratio. }
\end{table}

Among the analyzed countries in our study, only Vietnam belongs to Group 1. Our analysis revealed that while there are slightly more confirmed cases among men compared to women, i.e. $\rho_{cases} = 1.02$ in Vietnam, the male-to-female ratio of deaths is still smaller $\rho_{deaths} = 0.59$ and therefore more deaths were reported for women. Thus, this case seems to be contradictory to global data which indicates that men are more likely to get severely affected by COVID-19, and die more from the disease than women. One of the possible reasons is women’s representation in certain sectors strongly hit by the pandemic, such as the garment and textile sector, in some Asian countries including Vietnam, Cambodia and Sri Lanka. 
This can translate into two potential explanations motivating more deaths in women: 1) they become unemployed and without access to healthcare to deal with the disease, or 2) they become more vulnerable and most affected by cotton industry-related respiratory diseases related with the lack of safety equipment in unhygienic, unsafe environments with hazardous work conditions, as reported in \cite{kabir2019health,silpasuwan2016cotton}.
However, more research needs to be done in order to provide more insights on the vulnerability of women to COVID-19 in Vietnam.

\begin{figure}[htbp!]
    \centering
    \includegraphics[width=\textwidth]{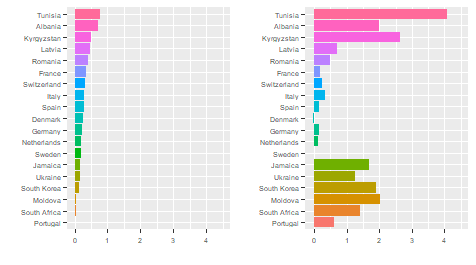}
    \caption{Log-scaled \textit {Male-to-Female} Deaths ratio (Left) vs Log-scaled \textit{Male-to-Female} smoking -female-to-male- ratio smoking (Right) for Group 2 (more cases in women but more deaths for men). This group is composed by 16 countries and shows that one possible explanatory variable is the factor \textit{smoking}, since men are shown to smoke more in these countries.}
    \label{fig:group2}
\end{figure}

\begin{figure}[htbp!]
    \centering
    \includegraphics[width=\textwidth]{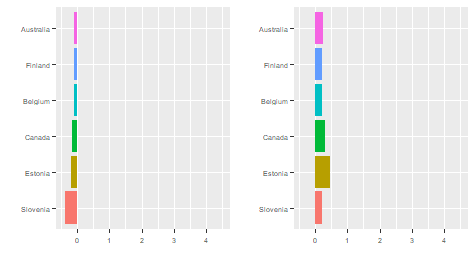}
    \caption{Log-scaled \textit{Male-to-Female} Deaths ratio (Left) vs Log-scaled \textit{Male-to-Female} Smoking ratio (Right) for Group 3 (6 countries, in which both cases and death ratios are higher for women, i.e., the opposite of most articles claims). In these countries, women smoke almost equally as men, and thus, smoking does not seem to clearly be an explanatory variable: women die as much or more than men.}
    \label{fig:group3}
\end{figure}

\begin{figure}[htbp!]
    \centering
    \includegraphics[width=\textwidth]{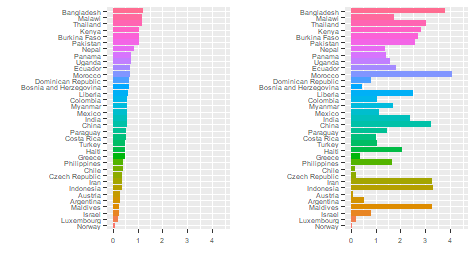}
    \caption{Log-scaled \textit{Male-to-Female} Deaths ratio (Left) vs Log-scaled \textit{Male-to-Female} Smoking ratio (Right) for Group 4 (35 countries where both cases and deaths are higher for men). This plot may reveal different testing strategies, as men are always more impacted.}
    \label{fig:group4}
\end{figure}

In Group 2, which contains 19 countries, women were more contaminated by COVID-19 than men. However, the number of deaths among male was higher. Data for this group also shows that this might be related to the much higher smoking rate in these countries. As shown for example in Fig. \ref{fig:group2}, a very high male-to-female smoking ratios are observed in most of the countries in this Group. Particularly, the highest smoking rates are observed in Tunisia, Albania and Kyrgyzstan, which also have the highest smoking ratios. Note that, in Figures \ref{fig:group2}, \ref{fig:group3}, \ref{fig:group4} we applied log scaling to the calculated ratios in order to plot them on a comparable scale. That is, a positive male-to-female smoking or death log-scaled ratio indicates a higher number of smoking or death among men, while a negative male-to-female smoking or death log-scaled ratio indicates a higher number of smoking or death among women.
While smoking might be one of the reasons that increases the risk of hospitalisation and death by COVID-19, as it is the case for most respiratory diseases, more data is needed in order to provide evidence on this hypothesis, such as age, number of tests by gender, etc.  

Driven by the observations we made in the previous group of countries, we were also interested in investigating the association between deaths ratios and smoking ratios for Group 3 and 4. Fig. \ref{fig:group3} and Fig. \ref{fig:group4} report the \textit{male-to-female} deaths ratio vs the \textit{male-to-female} smoking ratio for Group 3 and Group 4, respectively.
Group 3 represents 6 countries in which both confirmed and fatality rates are higher for women compared to men, while Group 4 represents 35 countries in which both confirmed cases and deaths are higher for men. Figures \ref{fig:group3} and \ref{fig:group4} also show a possible association between smoking and deaths. While \textit{male-to-female} log-scaled smoking ratios range up to $0.2$ across countries in Group 3, their values are much higher and range from $0.04$ up to $1.65$ in Group 4, in which the \textit{male-to-female} deaths ratios are also higher. It is also possible that countries in Group 3 are more likely to apply fairer testing strategies compared to the countries in Group 4, that have the highest \textit{male-to-female} death ratios. 

While our analysis suggests a possible association between smoking and a higher number of COVID-19 deaths, as most countries having a high \textit{male-to-female} deaths ratios, have a high \textit{male-to-female} smoking ratios as well, there is no firm conclusion that can be drawn regarding the relationship between smoking, 
sex and COVID-19. In addition, countries considered different criteria during the pandemic for reporting COVID-19 deaths and this could make understanding the impact of sex on COVID-19 ambiguous.

The differential findings and disparities observed across the four groups of countries in our analysis emphasize the need to understand why COVID-19 impacts some groups more than others. This might reflect other related factors and issues that need to be addressed, such as incomplete data and decision making biases.

These results motivated us to further study potential sources of bias that may have impacted the data analysis and thus, led us to analyze the kind of bias the data collection process may have suffered, as well as the fairness frameworks available to mitigate them. Since handling bias through fairness notions is often better understood through examples than definitions, next sections will illustrate both definitions, metrics and notions through use cases. First of all, next section is 
guided through a set of applied examples aimed at better conveying unintended consequences from data-driven analyses. 

\section{
Avoiding potential unintended discrimination policies} 
\label{sec:fair}
Different policies can be designed around pandemics such as COVID-19's regarding the outcome of different large-scale data analyses. For instance, policies on vaccination, free masks distribution, or assignment of preferential treatment based on the perception of vulnerability of a group. The availability of big data together with powerful statistical tools can lead to faster, evidence based solutions to pressing problems. However, it is imminent that data-driven decisions must both be accurate and fair. As it is defined by EU regulations (Art. 20, 21 EU Charter of Fundamental Rights, Art. 4 Directive 2004/113 and other directives) fair decision means no individual received less favorable treatment based on a value of a sensitive attribute (gender, race, ethnicity) \cite{ntoutsi2020bias}.

The issue of fairness in data-driven decisions, particularly, machine learning has been brought to the attention of the scientific community relatively recently \cite{zou2018ai,barocas2019fairness,verma2018fairness}, but its urgency is being continuously proved by examples of cases gone wrong \cite{hendricks2018women,viviano2021saliency,maguolo2020critic}. Here we showcase some examples, where biased AI decisions in business, justice and healthcare created disparate outcome \cite{zou2018ai} for disadvantaged groups: 
\begin{itemize}

\item \textit{Example 1}. A study published in Science 
found that a health care risk-prediction algorithm demonstrated racial bias because it relied on a race-correlated attribute. To compute who should qualify for the extra care, the algorithm designers used previous patients’ health care spending as a proxy for medical needs. Because black people were spending less money on healthcare, the program was less likely to flag eligible black patients for high-risk care management \cite{obermeyer_dissecting_2019}.

\item \textit{Example 2.} Similarly, the Amazon hiring case shows that training employment prediction algorithms on historical data, where women in technology have been a minority, will not make fair decisions towards female candidates in the future \cite{ajunwa2019paradox}. 

\item \textit{Example 3}. Another controversial case is using algorithmic tools to predict criminal behavior, and thus make decisions on prison sentences. In theory, this should reduce bias influencing the process, because judges would be making decisions based on data-driven recommendations. However, training algorithms on historic crime data puts groups historically disproportionately targeted by law enforcement --especially low-income and minority communities-- at a higher risk of being labeled as recidivists \cite{brackey2019analysis}. The inability to simultaneously satisfy all fairness criteria was demonstrated, showing that, in cases where recidivism prevalence differs across groups, disparate impact can arise 
when assessing the likelihood of recidivism \cite{chouldechova2017fair}.

\end{itemize}
In the above mentioned cases, bias arises from the data through the resurfacing of historic discrimination or spurious correlations that do not reflect causality \cite{pearl2018book}. 

Bias is particularly hard to control when inferences are made from observational data. Current COVID-19 research is mostly data-driven and based on observational data, aggregated from different sources. As a consequence, this opens many pathways for the results to be distorted. Specifically, an already widely accepted discovery of women being more resilient to the disease can be affected by spurious correlation, confounder and representation bias. 

Next, we illustrate unintended negative consequences for women if clinicians base decisions on assumptions of disparate vulnerability.

\subsection{
Illustrating the impact of using gender for COVID-19 treatment decisions 
within different causal contexts for confounder-based potential discrimination}


 One may argue that scientific freedom allows unconstrained search for knowledge. Research is evaluated based on its validity and robustness. However, the decisions in the real world based on biased data can create disparities in treatment or disparate impact resulting in disadvantage for protected groups. Disparate treatment is a variation in decisions for individuals that depend on the values of a sensitive attribute. 
 Disparate impact occurs when the decision outcomes disproportionately benefit or hurt members of certain sensitive attribute value groups \cite{zafar2017fairness}. 
 
 The adequate evaluation on fairness of decisions depends on the situation where the results of data analysis will be applied. Here we would like to illustrate the evaluation of fairness in the COVID-19 pandemic context, by modeling a situation where inference about women being less vulnerable to the virus is used for assigning a priority treatment to an individual (for example, a longer hospitalization, closer monitoring or access to vaccines). 
 
 We assume that higher vulnerability or risk of the severe symptoms for men is inferred from observing more cases of hospitalized individuals (variable $COVID$) in the electronic health records data. We also describe three hypothetical causal mechanisms that explain lesser female COVID-19 cases in the data. It should be noted that the causal graphs used to illustrate those mechanisms are not obtained by using a formal causal inference methods on the data as described by J. Pearl \cite{Pearl_review}, but are constructed as  visualizations for our described scenarios. 
 
 The three hypothetical confounders are $Exposure$, indicating if the individual is more exposed to the infection, for example by commuting to work or frequenting public places, $Smoking$, indicating whether the individual is a smoker, and $Hormones$ indicating gender dependent hormones. In the scenario where the confounding variable is $Exposure$, we consider a traditional society where women stay at home more than men and are less exposed to possibilities of being infected. In the second scenario we assume that on average women smoke less than men. In the third causal context female body chemistry is serving as protection from the severe cases of the virus, while male hormones increase mortality risk. All the three variables are responsible for association of 
 sex and number of registered cases in the hospitals, but themselves are not observed in the dataset. Therefore the only attribute that the decision can be based on is $Sex$. 
 
 We will illustrate the difference on fairness and the negative impact of classifying individuals for priority treatment based on $Sex$ under three scenarios considering those three confounders. We will consider only one confounding variable for each scenario for simplicity. In reality all three variables, as well as other health condition indicating features, can influence individual vulnerability to the virus. For example, even if female hormones would serve a protecting role, the mortality risk is likely to also be impacted by age, health status and cannot be inferred solely from 
 sex. In our purely hypothetical scenarios we analyze situations where sex could be considered as \textit{one of the variables predicting} vulnerability. In the context of our hypothetical scenarios we consider the use of the sex variable fair if the confounder has a causal link both with gender and vulnerability.

\textbf{Discrimination Scenario 1}: \textit{Negative impact on majority of women.} 

In this case the unobserved variable \emph{Exposure} indicates if a person gets more exposed to the virus, for example by commuting to work outside of home. \emph{Exposure} cannot be a resolving variable even if it is observed, because it is only associated with risk of being infected, but not the risk of mortality once already sick. However, it creates a statistical association between \emph{COVID} cases and \emph{Sex} through traditional gender roles. As a consequence, under this scenario, all women would be discriminated by being \textbf{systematically denied priority treatment} based on a false assumptions (such as those concluded by most papers in Table \ref{tab:covidPapers}). However, if we were predicting the likelihood of being infected AND the variable \emph{Exposure} was observed, then, this prediction could be used, for example, to distribute free masks to the more exposed individuals. On aggregate, men would get more masks if they are more likely to work outside of home. However, the unequal outcome would be resolved, because it is independent of sex or gender given that we know the individual is working and thus, has higher risk to get infected.

\textbf{Discrimination Scenario 2}: \textit{Negative impact on minority of women-smokers.} 

In this case the association between \emph{Sex} and \emph{COVID} cases is created by gender roles related confounding variable \emph{Smoking}, which could be a resolving variable, if it were observed. \emph{Smoking} has a causal effect on severeness of the disease, so assigning a priority \emph{Treatment} to smoking individuals is adequate. The disparity would be resolved even if more men than women will be assigned priority treatment, because on average men smoke more. Otherwise, if \emph{Smoking} is not observed, and the decision depends directly on \emph{Gender}, then the minority of women that are smokers will be \textbf{denied necessary priority treatment}.

\textbf{Discrimination Scenario 3}: \textit{Equal outcome for men and women.} 

Here we hypothesize that female hormones act as a protective factor, explaining the larger male vulnerability to COVID-19.
The resolving variable here is \emph{Hormones}, which is sex-specific. If expert consensus can confirm that the value of the resolving variable is consistent across all the individuals in a group (as a biologically predetermined attribute), then \textbf{the resolving variable is equivalent to the sensitive variable}. In this case, provided there are no other confounding variables, all men would be prioritized relative to the treatment received by women, but the disparity would be based on need and the outcome of this decision (and health impact) would be fair to women and men. On the other hand, if in this case we insisted for equal treatment, the men would suffer negative impact. 

\begin{landscape}
\begin{figure}[h]
\centering
 \begin{minipage}[t]{0.3\linewidth}
\resizebox{\linewidth}{!}
{
\begin{tikzpicture}
[
roundnode/.style={circle, draw=black!60, fill=white!5}
]
\node[roundnode, draw=blue!60, fill=blue!5,line width=1mm, minimum size= 2.5cm] (maintopic)                        {COVID-19 Cases};
\node[roundnode,draw=blue!60, fill=blue!5, line width=1mm, minimum size=2.5cm] (uppercicle) [above=of maintopic]  {Exposure};
\node[roundnode, dashed,draw=blue!60, fill=blue!5, line width=1mm, minimum size=2.5cm] (rightcicle) [right=3 cm of maintopic]  {Vulnerability};
\node[roundnode, draw=blue!60, fill=blue!5, line width=1mm, minimum size=2.5cm] (leftcircle) [left= 3 cm of maintopic]   {Sex};
\node[roundnode, draw=blue!60, fill=blue!5, line width=1mm, minimum size=2.5cm] (lowercircle) [below= 3 cm of maintopic] {Priority Treatment};

\draw[->,line width=1mm] (uppercicle.south) -- (maintopic.north)node [midway,above, fill=white]{Real cause of less cases for women};
\draw[->, dashed, line width=1mm] (maintopic.east) -- (rightcicle.west)node [midway,above, fill=white]{};
\draw[->, line width=1mm,dashed,  color=red] (leftcircle.east) -- (maintopic.west)node [midway,above, fill=white]{};
\draw[->, dashed, line width=1mm] (rightcicle.south) -- (lowercircle.east)node [midway,above, fill=white, anchor=center, below, text width=2.0cm]{Unfair decision};
\draw[->, dashed, line width=1mm] (maintopic.south) -- (lowercircle.north)node [midway,above, fill=white]{};


\draw[->, dashed, line width=1mm] (uppercicle.east)  -- (rightcicle.north)node [midway,above, fill=white]{};;

\draw[->, dashed, line width=1mm] (leftcircle.north) -- (uppercicle.west)node [midway,above, fill=white]{association};

\end{tikzpicture}
}
\label{fig:1}
\end{minipage}\hfill
\begin{minipage}[t]{0.3\linewidth}
\resizebox{\linewidth}{!}
{
\begin{tikzpicture}
[
roundnode/.style={circle, draw=black!60, fill=white!5}
]
\node[roundnode, draw=blue!60, fill=blue!5,line width=1mm, minimum size= 2.5cm] (maintopic)                        {COVID-19 Cases};
\node[roundnode,draw=blue!60, fill=blue!5, line width=1mm, minimum size=2.5cm] (uppercicle) [above= of maintopic]  {Smoking};
\node[roundnode, dashed,draw=blue!60, fill=blue!5, line width=1mm, minimum size=2.5cm] (rightcicle) [right=3 cm of maintopic]  {Vulnerability};
\node[roundnode, draw=blue!60, fill=blue!5, line width=1mm, minimum size=2.5cm] (leftcircle) [left= 3 cm of maintopic]   {Sex};
\node[roundnode, draw=blue!60, fill=blue!5, line width=1mm, minimum size=2.5cm] (lowercircle) [below= 3 cm of maintopic] {Priority Treatment};


\draw[->,line width=1mm] (uppercicle.south) -- (maintopic.north)node [midway,above, fill=white]{Real cause of less cases for women};
\draw[->, dashed, line width=1mm] (maintopic.east) -- (rightcicle.west)node [midway,above, fill=white]{};
\draw[->, dashed,line width=1mm, color = red] (leftcircle.east) -- (maintopic.west)node [midway,above, fill=white]{};
\draw[->, dashed,line width=1mm] (maintopic.south) -- (lowercircle.north)node [midway,above, fill=white]{};

\draw[->, line width=1mm] (uppercicle.east)  -- (rightcicle.north)node [midway,above, fill=white]{};
\draw[->, line width=1mm] (rightcicle.south) -- (lowercircle.east)node [midway,above, fill=white, anchor=center, below, text width=2.0cm]{Fair decision if based on smoking variable};
\draw[->, dashed,line width=1mm] (leftcircle.north) -- (uppercicle.west)node [midway,above, fill=white]{association};

\end{tikzpicture}
}


\label{fig:2}
\end{minipage}\hfill
\begin{minipage}[t]{0.3\linewidth}
\resizebox{\linewidth}{!}
{
\begin{tikzpicture}
[
roundnode/.style={circle, draw=black!60, fill=white!5}
]
\node[roundnode, draw=blue!60, fill=blue!5,line width=1mm, minimum size= 2.5cm] (maintopic)                        {COVID-19 Cases};
\node[roundnode,draw=blue!60, fill=blue!5, line width=1mm, minimum size=2.5cm] (uppercicle) [above=of maintopic]  {Hormones};
\node[roundnode, dashed, draw=blue!60, fill=blue!5, line width=1mm, minimum size=2.5cm] (rightcicle) [right=3 cm of maintopic]  {Vulnerability};
\node[roundnode, draw=blue!60, fill=blue!5, line width=1mm, minimum size=2.5cm] (leftcircle) [left= 3 cm of maintopic]   {Sex};
\node[roundnode, draw=blue!60, fill=blue!5, line width=1mm, minimum size=2.5cm] (lowercircle) [below= 3 cm of maintopic] {Priority Treatment};

\draw[->,line width=1mm] (uppercicle.south) -- (maintopic.north)node [midway,above, fill=white]{Real cause of less cases for women};
\draw[->, dashed, line width=1mm] (maintopic.east) -- (rightcicle.west)node [midway,above, fill=white]{};
\draw[->, dashed,line width=1mm, color=green] (leftcircle.east) -- (maintopic.west)node [midway,above, fill=white]{};
\draw[->, dashed,line width=1mm] (maintopic.south) -- (lowercircle.north)node [midway,above, fill=white]{};

\draw[->, line width=1mm] (uppercicle.east)  -- (rightcicle.north)node [midway,above, fill=white]{};
\draw[->, line width=1mm] (rightcicle.south) -- (lowercircle.east)node [midway,above, fill=white, anchor=center, below, text width=2.0cm]{Fair decision};
\draw[->, line width=1mm] (leftcircle.north) -- (uppercicle.west)node [midway,above, fill=white]{cause};

\end{tikzpicture}
}
\label{fig:3}
\end{minipage}
\caption{The schemes illustrate three scenarios and corresponding biases: First (left) confounder (exposure) bias creates association between gender and vulnerability. Second (middle) confounder (smoking) bias leads to wrong inference about women being less vulnerable. Third (right) case: no bias given no other confounders are controlled or not present. Solid black lines represent causal links and dashed lines stand for associations. The path leading to \textit{treatment} decision is considered fair if it goes through causal links, and therefore no negative impact is created. Red and green colors indicate if a gender/sex variable can at all be considered in the prediction of vulnerability. The only "green" case is the one where a causal link connects sex and a vulnerability factor.} 
\end{figure}
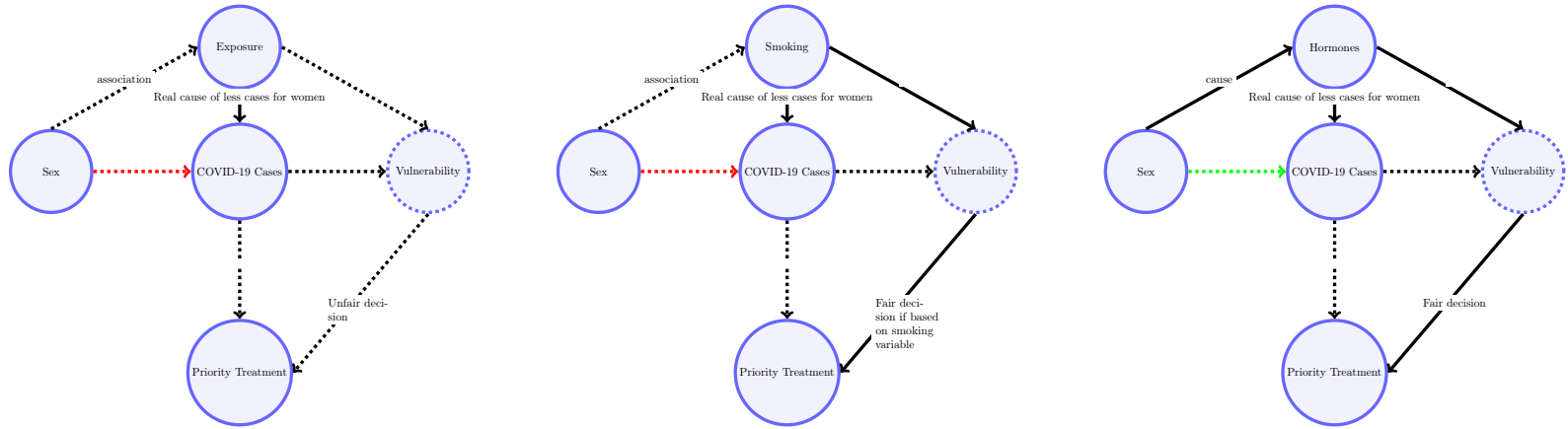
\label{fig:graphs}
\end{landscape}
In contrast to our simplified scenarios, in reality a set of biases can be entangled to play all in favour of discrimination based on false assumptions of vulnerability and due to lack of information. Biases may support and cover each other \cite{cofone2018algorithmic}. The complexity of relationships and unequal fairness consequences depending on them implies a necessity of human experts in the loop and therefore, a model that is not only explainable but also interactive. 
More complex graphs faithful to the causal relationships learned from the data can be a useful tool to ensure explainability and evaluate causability\footnote{\textit{Explainability}  highlights technically relevant parts of
machine representations and machine models; \textit{Causability} 
measures the extent to which an explanation to a human achieves a specified level of causal understanding \cite{holzinger_causability_2019}.} and fairness of the subsequent decision policies.

Since it may be hard to disentangle when more than one bias is taking place, the following section will aim at clarify the overall categories or dimensions to pay attention to, when control resources are urgent or limited.

\section{Detecting and handling Bias: \textit{The Bias Priority 
Recommendations} (BPR) and the \textit{Bias Catalogue for Pandemics} }
\label{sec:bias_priority}

After identifying the large amount of sources of bias a 
large-scale data-driven analysis can suffer, we deem relevant listing the set of potential bias a research study is vulnerable to. In order to follow a scale of priority when designing a study or its ethical approval, we defined a taxonomy of large bias families to attend to. 

Next we present these 
identified subdivisions of bias, its concrete types, and its urgency of being examined and tackled 
when dealing with pandemics and disaster protocols requiring rapid response.

Inspired by the software engineering MoSCoW methodology (see more in the Appendix), we established a 2-level priority scale  (P1, P2) on which to assess each type of bias given a particular research question to be addressed by large scale observational data analysis. In order to group and prioritize the set of biases to study, the \textit{Bias Priority Recommendations} we propose are based on the following criteria: 
\begin{itemize}
    \item Frequency of occurrence of bias 
    \item Ubiquity of bias
    \item Level of impact of bias
    \item Early introduction of the bias in the data preprocessing pipeline
    \item Testing easiness of bias (how easy is to check that the bias occurs).
\end{itemize}



Tables \ref{tab:priorities1} and \ref{tab:priorities2} show how we categorized the potential biases included in the Bias Catalog for Pandemics into a scale of priorities according to decreasing urgency (P1, P2). 
The aim is to facilitate task prioritization in projects when facing scarcity in time, funding, or human and technical resources. 


\begin{landscape}

\begin{table}[htbp!]

 \begin{center}
 \centering
 \begin{adjustbox}{width=2\textwidth}
 \begin{tabular}{ | c | c | c | c |}
 \hline
\textbf{Bias Family} & \textbf{Bias Name} & \textbf{Bias} & \textbf{Key point} \\
\textbf{} & \textbf{} & \textbf{Priority} & \textbf{} \\
\hline

\multirow{6}{*}
{\vtop{\hbox{\strut Sampling Biases}}}

& \multicolumn{1}{l|}{Diagnostic access Bias} & 
   \multicolumn{1}{l|}{P1} &
    \multicolumn{1}{l|}{\parbox{14cm}{Different geographic, temporal, and economic access affect to diagnostic procedures for a given disease}}\\\cline{2-4}
    \\[-1em]
    
& \multicolumn{1}{l|}{Volunteer Bias} & 
   \multicolumn{1}{l|}{P2} &
    \multicolumn{1}{l|}{\parbox{14cm}{Participants volunteering to take part in a study intrinsically have different characteristics from the general population of interest}}\\\cline{2-4}
    \\[-1em]
    
& \multicolumn{1}{l|}{Information Bias} & 
   \multicolumn{1}{l|}{P1} &
    \multicolumn{1}{l|}{\parbox{14cm}{Systematic differences 
    in the collection, recall, recording or handling of information}}\\\cline{2-4}
     \\[-1em]
     
& \multicolumn{1}{l|}{Informed presence Bias} & 
   \multicolumn{1}{l|}{P2} &
    \multicolumn{1}{l|}{\parbox{14cm}{
    Electronic health records contain information about specific parts of the population which are different than the general population, which can possibly create spurious associations}}\\\cline{2-4}
     \\[-1em]

& \multicolumn{1}{l|}{Availability Bias} & 
   \multicolumn{1}{l|}{P2} &
    \multicolumn{1}{l|}{A distortion arising from the information which is most readily available}\\\cline{2-4}
              \\[-1em]

& \multicolumn{1}{l|}{Unacceptability Bias} & 
   \multicolumn{1}{l|}{P2} &
    \multicolumn{1}{l|}{\parbox{14cm}{A systematic difference in response rates of tests due to their “unacceptability”}}\\\cline{2-4}
     \\[-1em]
     
& \multicolumn{1}{l|}{Unacceptable disease Bias} &                                 \multicolumn{1}{l|}{P2} &
    \multicolumn{1}{l|}{\parbox{14cm}{Lower rates of reporting of certain \textit{unacceptable} diseases}} \\
    \hline
     \\[-1em]
     
& \multicolumn{1}{l|}{Prevalence-incidence (Neyman) Bias}&  
   \multicolumn{1}{l|}{P2} &
    \multicolumn{1}{l|}{\parbox{14cm}{Exclusion of individuals with severe or mild disease resulting in a systematic
error in the estimated association or effect of an exposure on an outcome}}\\\cline{2-4}
     \\[-1em]
     
& \multicolumn{1}{l|}{Misclassification Bias} & 
   \multicolumn{1}{l|}{P2} &
    \multicolumn{1}{l|}{\parbox{14cm}{A study participant being categorised into an incorrect category}}\\\cline{2-4}
     \\[-1em]
     
& \multicolumn{1}{l|}{Compliance Bias} & 
   \multicolumn{1}{l|}{P2} &
    \multicolumn{1}{l|}{\parbox{14cm}{Participants compliant with an intervention differ in some way from those
not compliant}}\\\cline{2-4}
 \\[-1em]
 
& \multicolumn{1}{l|}{Selection Bias} & 
   \multicolumn{1}{l|}{P1} &
    \multicolumn{1}{l|}{\parbox{14cm}{A different selection procedure leads to a systematic error in an outcome}}\\\cline{2-4}
     \\[-1em]

\multirow{16}{*}
{\vtop{\hbox{\strut Diagnostic Biases}}}

& \multicolumn{1}{l|}{Verification Bias} & 
   \multicolumn{1}{l|}{P2} &
    \multicolumn{1}{l|}{\parbox{14cm}{When only a proportion of the study group receives confirmation of the diagnosis by the reference standard, or if some patients receive a different
reference standard at the time of diagnosis}}\\\cline{2-4}
 
     \\[-1em]  
     
& \multicolumn{1}{l|}{Differential reference Bias} & 
   \multicolumn{1}{l|}{P1} &
    \multicolumn{1}{l|}{\parbox{14cm}{Lack of receiving the same reference test in a diagnostic accuracy study}}\\\cline{2-4}
     \\[-1em]
     
& \multicolumn{1}{l|}{Partial reference Bias} &                                  \multicolumn{1}{l|}{P1} &
    \multicolumn{1}{l|}{\parbox{14cm}{Verification bias %
    overestimating sensitivity and specificity 
    of a new test against a reference standard test when 
    the reference one is invasive, expensive or carries risk. }} \\\cline{2-4}
 \\[-1em]
& \multicolumn{1}{l|}{Mimicry Bias} & 
   \multicolumn{1}{l|}{P2} &
    \multicolumn{1}{l|}{\parbox{14cm}{Occurs when an innocent exposure causes a benign disorder that resembles the disease. 
    }}\\\cline{2-4}
     \\[-1em]
& \multicolumn{1}{l|}{Substitution game Bias} & 
   \multicolumn{1}{l|}{P2} &
    \multicolumn{1}{l|}{\parbox{14cm}{When substituting a clinically important endpoint or an exposure with a surrogate marker for the disease in the absence of data}}\\\cline{2-4}
     \\[-1em]
     
& \multicolumn{1}{l|}{Unmasking (detection signal) Bias} &           \multicolumn{1}{l|}{P2} &
    \multicolumn{1}{l|}{\parbox{14cm}{An innocent exposure causing a sign or symptom that precipitates a search for the disease}} \\\cline{2-4}
 \\[-1em]
 
& \multicolumn{1}{l|}{Spectrum Bias} & 
   \multicolumn{1}{l|}{P2} &
    \multicolumn{1}{l|}{\parbox{14cm}{The sensitivity and specificity of diagnostic tests vary in different way}}\\\cline{2-4}
     \\[-1em]
    
& \multicolumn{1}{l|}{Detection Bias} & 
   \multicolumn{1}{l|}{P2} &
    \multicolumn{1}{l|}{\parbox{14cm}{Systematic differences between groups in how outcomes are determined}}\\\cline{2-4}
     \\[-1em]
    
& \multicolumn{1}{l|}{Insensitive measure Bias} & 
   \multicolumn{1}{l|}{P1} &
    \multicolumn{1}{l|}{\parbox{14cm}{When the tool or test outcome is not accurate}}\\\cline{2-4}
     \\[-1em]
& \multicolumn{1}{l|}{Incorporation Bias} & 
   \multicolumn{1}{l|}{P2} &
    \multicolumn{1}{l|}{\parbox{14cm}{Results of an index test form part of the reference test in a diagnostic study}}\\\cline{2-4}
     \\[-1em]

& \multicolumn{1}{l|}{Referral filter Bias} & 
   \multicolumn{1}{l|}{P2} &
    \multicolumn{1}{l|}{\parbox{14cm}{
    Participants may not properly represent the population 
    due to non applicability 
    }}\\\hline

\end{tabular}
\end{adjustbox}
\end{center}
\caption{Identified kind of biases and their priorities (P1- Must 
test, P2- Should test) for disaster response data-driven decisions (Part I).  Priorities are based on the frequency of occurrence of the bias, its potential impact on later decision making, and its appearance being earlier within the research/data science pipeline --which can lead to a domino effect later involving other biases--. 
The Bias family column includes six major kinds of bias categories identified as main dimensions to account for. For more details on each bias description we refer the reader to \cite{catalogOfBias17}.}
\label{tab:priorities1}
\end{table}

\end{landscape}


\begin{landscape}
\begin{table}[htbp!]
 \begin{center}
 \centering
\begin{adjustbox}{width=2\textwidth}
 \begin{tabular}{ | c | c | c | c |}
 \hline
\textbf{Bias Family} & \textbf{Bias Name} & \textbf{Bias} & \textbf{Key point} \\
\textbf{} & \textbf{} & \textbf{Priority} & \textbf{} \\
\hline

\multirow{11}{*}
{\vtop{\hbox{\strut Interpretation Biases}}}

     & \multicolumn{1}{l|}{Spin Bias} & 
   \multicolumn{1}{l|}{P2} &
    \multicolumn{1}{l|}{\parbox{14cm}{A distorted interpretation of research results misleads conclusions}}\\\cline{2-4} 
     \\[-1em]

     & \multicolumn{1}{l|}{Confounding Bias} & 
   \multicolumn{1}{l|}{P1} &
    \multicolumn{1}{l|}{\parbox{14cm}{A distorted association between exposure and outcome due to a factor independently associated with both}}\\\cline{2-4} 
     \\[-1em]             
              
& \multicolumn{1}{l|}{Data dredging Bias} & 
   \multicolumn{1}{l|}{P2} &
    \multicolumn{1}{l|}{\parbox{14cm}{A distortion arising from the results of statistical tests}}\\\cline{2-4}
     \\[-1em]

& \multicolumn{1}{l|}{Performance/ Lack of blinding Bias} & 
   \multicolumn{1}{l|}{P2} &
    \multicolumn{1}{l|}{\parbox{14cm}{The lack of adequate blinding in a trial results in systematic differences
    }}\\\cline{2-4}
     \\[-1em]
     
    
& \multicolumn{1}{l|}{Observer Bias} & 
   \multicolumn{1}{l|}{P1} &
    \multicolumn{1}{l|}{\parbox{14cm}{Observing and recording information creates systematic discrepancies from the truth}}\\\cline{2-4}
 \\[-1em]
 
& \multicolumn{1}{l|}{Perception Bias} & 
   \multicolumn{1}{l|}{P2} &
    \multicolumn{1}{l|}{\parbox{14cm}{When there is a tendency to be subjective about people and events, causing biased interpretation of a study or biased information collection}}\\\hline
 \\[-1em]

\multirow{2}{*}
{\vtop{\hbox{\strut Prior Belief Biases}}}  

& \multicolumn{1}{l|}{Ascertainment Bias} & 
   \multicolumn{1}{l|}{P2} &
    \multicolumn{1}{l|}{\parbox{14cm}{When there is more intense surveillance or screening for the outcome
of interest among exposed populations than among unexposed populations (also a \textit{Sampling Bias})}}\\\cline{2-4}
     \\[-1em]  
     
& \multicolumn{1}{l|}{Diagnostic suspicion Bias} & 
   \multicolumn{1}{l|}{P1} &
    \multicolumn{1}{l|}{\parbox{14cm}{Prior beliefs
about the likeliness of the disease held by patients or doctors affect the process and outcome of the diagnostic test outcomes}}\\\cline{2-4}
 \\[-1em]
& \multicolumn{1}{l|}{Previous opinion Bias} 

& \multicolumn{1}{l|}{P2} &
    \multicolumn{1}{l|}{\parbox{14cm}{Previous assessments 
    affect results of subsequent processes on the same patient}}
    \\\hline

\multirow{5}{*}
{\vtop{\hbox{\strut Research Aspirations Biases}}} 

& \multicolumn{1}{l|}{One sided reference Bias} &                                  \multicolumn{1}{l|}{P2} &
    \multicolumn{1}{l|}{\parbox{14cm}{Authors restricting their references to 
    those supporting their position}} \\\cline{2-4}
     \\[-1em]

& \multicolumn{1}{l|}{Reporting Bias} &                                            \multicolumn{1}{l|}{P2} &
    \multicolumn{1}{l|}{\parbox{14cm}{Selective revelation or suppression of information or study results}} \\\cline{2-4}
    
 \\[-1em]    

& \multicolumn{1}{l|}{Popularity Bias} & 
   \multicolumn{1}{l|}{P1} &
    \multicolumn{1}{l|}{\parbox{14cm}{The uptake of healthcare 
    public interest in a particular disease causes biased results}}\\\cline{2-4}
     \\[-1em]
                       
& \multicolumn{1}{l|}{Positive results Bias} & 
   \multicolumn{1}{l|}{P2} &
    \multicolumn{1}{l|}{\parbox{14cm}{The tendency to submit, accept and publish positive results rather than negative ones}}\\\cline{2-4}
     \\[-1em]

& \multicolumn{1}{l|}{Outcome Reporting Bias} & 
   \multicolumn{1}{l|}{P1} &
    \multicolumn{1}{l|}{\parbox{14cm}{Occurs when there is a selective reporting of pre-specified outcomes in published clinical trials}}\\\cline{2-4}
     \\[-1em]
     
& \multicolumn{1}{l|}{Hot stuff Bias} & 
   \multicolumn{1}{l|}{P2} &
    \multicolumn{1}{l|}{\parbox{14cm}{When a research topic is perceived to be of
imminent relevance or have a controversial aspect}}\\\cline{2-4}
 \\[-1em]

& \multicolumn{1}{l|}{Novelty Bias} & 
   \multicolumn{1}{l|}{P2} &
    \multicolumn{1}{l|}{\parbox{14cm}{The tendency for an intervention to appear better when it is new (Also a \textit{Timing Bias})}}
    \\\hline
     \\[-1em]

\multirow{3}{*}
{\vtop{\hbox{\strut Timing Biases}}}
& \multicolumn{1}{l|}{Non-contemporaneous control Bias} 
&   \multicolumn{1}{l|}{P2} &
    \multicolumn{1}{l|}{\parbox{14cm}{Timing differences when selecting case and controls influence outcomes}}\\\cline{2-4}
     \\[-1em]
     
& \multicolumn{1}{l|}{Chronological Bias} &                             \multicolumn{1}{l|}{P2} &
    \multicolumn{1}{l|}{\parbox{14cm}{participants allocated earlier to an intervention or a group are at a different risk from participants who are recruited at a later stage}} \\\cline{2-4}
     \\[-1em]
    
& \multicolumn{1}{l|}{Starting time Bias} & 
   \multicolumn{1}{l|}{P2} &
    \multicolumn{1}{l|}{\parbox{14cm}{Failure to identify a common starting time for a disease}}\\\cline{2-4}
     \\[-1em]
& \multicolumn{1}{l|}{Immortal time Bias} & 
   \multicolumn{1}{l|}{P2} &
    \multicolumn{1}{l|}{\parbox{14cm}{Participants in the exposed group cannot experience the outcome}}
     \\\hline

 \end{tabular}
 \end{adjustbox}
 \end{center}
\label{tab:priorities}
\caption{Identified kind of biases and their priorities (P1- Must 
test, P2- Should test) for disaster response data-driven decisions (Part II).  Priorities are based on the frequency of occurrence of the bias, its potential impact on later decision making, and its appearance being earlier within the research/data science pipeline --which can lead to a domino effect later involving other biases--. 
The Bias family column includes six major kinds of bias categories identified as main dimensions to account for. For more details on each bias description we refer the reader to \cite{catalogOfBias17}.}
\label{tab:priorities2}
\end{table}
\end{landscape}

We 
relied on the Oxford Catalogue of Bias Collaboration\footnote{\url{https://catalogofbias.org/}} 
\cite{catalogOfBias17} in order to illustrate potential problems of the studies exposed earlier regarding COVID-19-related claims. We exclusively focus on biases prone to affect COVID-19 observational studies that can potentially evolve into discriminatory policy making. We propose solutions to the data-driven biases and provide a recommendations list to serve as guidelines to combat them. 
As the literature shows, a large amount (50 in \cite{catalogOfBias17}) of identified biases were considered in our initial screening for biases. We later filtered to those applicable to the sex and gender COVID-19 pandemic context. Out of the 50, 38 are directly or indirectly applicable to the analysed literature on the impact of gender on COVID-19 and/or pandemics in a generic sense. 

Our contributed full \textit{Bias Catalogue for Pandemics} further expanded with focus on COVID-19 contextual examples (in the appendix), includes examples, potential solutions and preventive steps in pandemics-like situations. 
As a result of the compendium of biases to be aware of, and the claims reported in the latest literature, we are able to draw the recommendations below.\newline \newline

\textbf{The Bias Priority Recommendations (BPR)}:

\begin{rec}
\textit{Planning for additional data collection}. \end{rec}
Since more data needs to be gathered to contextualize studies and perform post-hoc bias analysis, collecting additional potential explainable variables could be the first step to find more confounding variables that explain the main differences in reported outcomes.

\begin{rec}\textit{Applying tests for each bias family}. \end{rec}
Assessing if at least one bias type from each defined bias family, specially those easier to test or of priority 1, is paramount in any data analysis prone to generate claims and, in turn, policies deriving on vulnerability of a group.

\begin{rec} \textit{Providing a \textit{Tested Biases List}}.\end{rec}
    Since a 100\% bias-free model is impossible \cite{chouldechova2017fair,maxwell2020facial,kilbertus2020fair}, we encourage to report the biases a) Accounted for, b) Tested, and c) Mitigated in each study or model outcome. 

\begin{rec}
\textit{Providing a \textit{Bias Disclaimer List} } 
\end{rec}
    Every study must provide a Disclaimer List of (un/)tested biases, to facilitate follow-up works by the scientific community

\begin{rec}
\textit{Placing and assessing studies within a 
Fairness Framework}
\end{rec}
    The selection of (observational or algorithmic) fairness frameworks is recommended, while choosing the one most adapted to the conditions, relevancy, frequency and potential impact of the study. For instance, fairness notions such as the \textit{demographic parity}\footnote{Demographic parity states that the output of the machine learning model should be equal between two or more groups. The demographic parity difference is then a measure of how much disparity there is between model outcomes in two groups of samples \cite{lundberg2017unified}.} would force the outcome to be equal between the groups, which may be not applicable in medical domains. The more nuanced causal fairness notions, for example, \textit{No unresolved discrimination}, can help evaluate the causal path leading from the sensitive variable to the decision \cite{kilbertus2018avoiding}.

\begin{rec}
\textit{Providing analyses and estimation of causal effects} 
\end{rec}
Approaching observational studies or large-scale testing regimes in risk-structured population models should assess causation or mitigation of effects of confounding or selection bias.

In observational studies performed in social media --often used when gold standard randomized controlled experiments are impractical--, methods that help assessing causation or mitigating the effects of confounding or selection bias are recommended as per \cite{olteanu2019social}\footnote{E.g., matched analysis, instrumental variables analysis, regression discontinuities, differences-in-differences, and other causal inference methods with observational data \cite{nichols2007causal}.}. 

In large scale testing regimes, for instance, causal graphs, risk-based testing frameworks, or validating testing models through death predictions --which are often assumed to be reported more accurately than the number of infections-- can be used \cite{besserve2021assaying}.

\begin{rec} 
\textit{Applying responsible AI principles and eXplainable AI techniques (XAI) \cite{arrieta2020explainable} to analyze feature contributions, and provide explainable, trustworthy and actionable model outputs}. 
\end{rec}
 The severity of the impact of inaccurate or unfair decisions implies ensuring the human supervision of AI model decisions and therefore, their explainability, causality and causability. When using powerful AI models that are black boxes difficult to interpret, explainability techniques can be used (e.g., \textit{contrastive} and \textit{counterfactual} explanations  \cite{stepin2021survey}). 
 Given the complexity of the relationship between factors, more outreaching causal, multimodal explainability and causability methods (that account for the user's acceptation of the explanation \cite{holzinger_2021}) should be employed. 

\begin{rec} \textit{Considering the learning of models that maximize utility under fairness constraints and state the accuracy-fairness trade-off}. 

For instance, assuring that a model takes fair decisions despite these being imperfect can display favorable properties in terms of the trade-off between utility and fairness. 
Learning stochastic decision policies in such way \cite{kilbertus2020fair} can shift the paradigm \textit{from “learning to predict” to “learning to decide”}. 

When using models under fairness constraints, it is also recommendable to study and state the joint optimization 
boundaries in the accuracy-fairness trade-off, which can also ideally include the model complexity as a third objective to minimize \cite{valdivia2020fair}.
\end{rec}

\begin{rec}
Framing research studies within existing bias and pitfall detection frameworks.

These help enlighten potential sources of bias at both the data sourcing and rest of data processing stages, pointing at methodological limitations and ethical boundaries.

In COVID-19-like pandemic scenarios we recommend following the CLAIRE Initiative COVID-19 recommendations \cite{bontempi2021claire} to facilitate data-driven AI solutions. 

For bias handling in social media data, see the guidelines in \cite{olteanu2019social}.
\end{rec}


\begin{rec}

Designing a common data platform 
with a shared \textit{cross-domain lingo} for bias identification and diagnosis.

The platform should contain tools that facilitate reproducibility and repeatability of results, as well as easy access and continuous maintenance for improved trust\footnote{\url{https://theconversation.com/covid-19-la-malinterpretacion-de-los-datos-de-la-pandemia-dana-la-confianza-del-publico-149387}}.

Before performing a clinical research study, this framework must be shared, understood and agreed upon among clinicians, physicians, epidemiologists, bio-statisticians and data scientists. As recommended by \cite{bontempi2021claire}, it should including open data space, infrastructure, open licenses, data management plan, and data standardization scheme.

\end{rec}

 


\section{Discussion}%
\label{sec:discussion}

Despite the amount of hospitalized males being larger in comparison with women, there may be several explanatory variables and behaviours. For instance, men lifestyle is different from women's, which can be a reason that men are more affected by COVID-19 infection. Men are more inclined towards drinking and smoking, which can evolve into lung infection which in turn, can formulate more chance of COVID-19 infection. 

All over the world, countries considered different criteria during the pandemic for reporting COVID-19 deaths, and this could explain the inequalities and potential disparate impact on 
sex and gender on COVID-19. Furthermore, each country is using several testing strategies and numerous testing methods are unreliable. Many lower-income countries have limited medical equipment. In addition, they are required to create biomarkers and deal with proper patient management systems in order to overcome COVID-19.

Social and cultural differences are additionally affecting the COVID-19 pandemic. 
In this line, another potential factor is the tendency of females to comply more with regulation, protecting themselves more and wearing masks more, and thus, these variables could explain such claims. 

Many countries in the world suffer gender disparity on the job market which, as a consequence, makes men more exposed and thus, more affected by the virus compared to women.  
In most countries, women bear the bulk of domestic responsibilities, including not only household chores but also care (notably for children) and, importantly, medical care. Women are typically in charge of ensuring health for the whole family as part of their traditional reproductive work. Their greater compliance with COVID-19 recommendations is a reflection of long-established gender social roles, which has also involved an increased burden for them during the pandemic \cite{Power2020}. 

Socio-cultural dimensions may reversibly affect the undertaking of preventive measures \cite{haischer2020wearing} and cyclic lockdown policies \cite{bwire2020coronavirus}. 
For instance, applications using the latest techniques in AI could aid to automatically detect mask wearing and hygiene practices from images in public places, e.g., supermarkets \cite{haischer2020wearing}. 
This is just an example among the many possible follow up studies we encourage in order to validate the potential explanations for these claims. 

Since the latest ML models such as deep networks do not correct against, but rather replicate existing biases of the researchers who train them, the data they are fed, the circumstances of their testing, etc., we hope more effort is initially put into both performing fair data collection and data analyses. Likewise these checks need to be present when developing methods able to programmatically verify, flag, and reduce data and model biases. Stating the verified tests and/or using our recommended protocols will minimally set the state of affairs on the table, and therefore, highlight and make legal processes stand up for process automation. As a positive side effect, AI-based accountability will be more easily gained. Fair data analysis is only the first step towards human-centric 
societies endowed with responsible AI systems that serve citizens and governments make use of data-informed policies more efficiently.


We encourage the branch of research that further develops the interest for integrating additional features that can reflect and potentially explain often unethical confounders. Future work should perform clinical trials with randomised assignment on different factors not studied in depth, i.e., 
hormonal treatment or different work-related exposure\footnote{\url{https://elpais.com/ciencia/2021-03-05/no-es-el-sitio-es-lo-que-ocurre-dentro-por-que-los-bares-y-los-restaurantes-suponen-riesgo-de-contagio.html}}. 
Especially supporting the last factor is the work published in Nature by \cite{chang2021mobility}, based on a mobile phone mobility study. It alerted of the role of hostelry in the infections, being restaurants the most dangerous, up to 3 or 4 times more than the following categories exposed: gyms, bars \cite{methi2021covid,fitzgerald2021managing} and hotels. While the tourism sector has been studied in  \cite{chica2021collective}, the labour market, skills and wage levels lockdown measures across sectors was reported in \cite{fana2020covid}. These are part of the factors to further study to have more fine grained informative claims.

\section{Conclusions and Future Work }
\label{sec:conclusions}

The different impact of male and female sex on COVID-19 implies the existence of diversity in the different biology (dependence on the biological immune system), health status, mortality rates, lifestyle (smoking), responsibilities, and others. 
Our analysis suggests a possible association between smoking and a higher amount of COVID-19 deaths, or a lack of healthcare support, 
as examples of a larger set of hypotheses to be studied.  
However, the exact mechanism contributing to this different outcome remains elusive and needs further investigation.

In addition, other explanations having more complex dimensions, and thus broader implications, include specific inequalities in gendered working forces characterized strongly by being associated to a particular gender, such as the garment industry in Asia. These could facilitate the generation of hypotheses linked to policies that can affect regions differently (e.g., see the supply chain ripple effect and lack of healthcare access implications due to unemployment on the garment industry in Asia and the Pacific \cite{international2020supplyChainRippleEffect,kabir2019health, silpasuwan2016cotton}, or the respiratory and other health vulnerabilities related to the garment industry workers).



Furthermore, a responsible attitude from females towards COVID-19 prevention \cite{haischer2020wearing} is another major element considered. More research is required to examine the influence of age and gender on COVID-19. To develop a better understanding of the true biological differences in disease propagation and adverse outcomes, more research is required in hormonal, inflammatory, immunologic, and phenotypical dimensions in severe COVID-19 disease.

While in this article we treated the sex and gender variables, the consequences of disparate impact can apply to any protected variable, and they can occur in any other analysis where data collection may suffer from bias. We dig further in the subject of COVID-19 and showed that the vulnerability of women toward COVID-19 is not really well justified when looking closer at studies across publications and merged databases. For instance, smoking seems to be a potential explanatory variable to study further, in the same way as Vitamin D deficit, or other yet to be found variables. However, our message, crystallized through the recommendation guidelines, is generic and therefore, orthogonal to our COVID-19 case study.  

The literature review and illustrated data analysis led us to observe that there are a large amount of tests for mitigating bias in large-scale data-driven analyses. 
Caution taken to avoid cherry-picking of results should include the use of available fairness frameworks and tests for biases, and always informing of the potential bias a study may suffer from.

We identified a set of priorities in the illustrated Bias Catalog for Pandemics, where we prioritized based on the frequency of occurrence of the bias, its potential impact on decision making and results, and its appearance at earlier stages of the full research process. This is because from a data science perspective, a bias that occurs in the collecting or preprocessing stages usually leads to a domino effect and introduces bias in the whole pipeline. 

Among over 50 kind of bias considered,  we identified 38 as potentially applicable to contexts such as our case study on the COVID-19 pandemic. We studied these as well to serve as an exercise, and find out which factors most likely can explain the differences in gender results. Interactions among the biases listed on our BCP are prone to explain any kind of disproportionate or non representative findings, and we invite the research community to become not only aware of, but also acquainted with them, as a go-to encyclopedia, and carefully use them, whenever possible, as guiding protocol for reporting aggregated results. 

The minimal set of bias identification tools and Bias Priority Recommendations (BPR) presented aim at aiding the design and reporting of research dealing with urgent socio-economic needs of the scale of pandemics. It is in these cases of world-wide affecting threats such as the COVID-19 pandemic that data can aid the most, but also have unintended and unexpected aggravating policies. Therefore, reporting must be done with caution. 

We hope this is the beginning of more methodically structured studies that consider fairness and bias from the very first data collection to the last analysis and so, lead to more curated claims that can avoid exposing minorities at risk.



 \section{Acknowledgement(s)}
We thank 
Songul Tolan and Golnoosh Farnadi 
for his helpful feedback. We also acknowledge Carlos Castillo for early feedback at the conceptualisation phase of the paper. 

\section{Appendix}
The first Appendix section further develops the compiled Bias Catalog \cite{catalogOfBias17} for Pandemics. The second section presents background on the source of the prioritizing scheme idea we followed to assign priorities to biases.

The Bias Catalog for Pandemics specifically extends the Bias Catalog \cite{catalogOfBias17} to attend to pandemic-like contexts, and enhances it with a set of COVID-19 pandemic examples, and a compilation of mitigating  policies or preventive steps as potential solutions.


\subsection{\textit{The Bias Catalog for Pandemics}: Sources of bias to account for when deriving data-driven claims }


 
Below we compile the Bias Catalog for Pandemics, which extends a collaborative project mapping all biases that affect health evidence \cite{sackett1979bias,catalogOfBias17} on cases specifically related to the topic of our paper, assessing the impact of sex and gender on COVID-19-like 
pandemics.  
\begin{mydef}
\textit{Ascertainment bias}\end{mydef}

Ascertainment bias generally refers to situations when data is collected in such a way to include some members of a population more than others. This can happen when there is more intense surveillance or screening for the outcome of interest among exposed populations than among unexposed populations.

* \textit{COVID-19 Context:}  The possibility of easy access for testing in some countries for men or for women could lead to this bias. One might find a higher rate of confirmed cases in the population with easy access to testing. The risk of death among confirmed cases in some populations could be therefore overestimated.

* \textit{Preventive steps:} Appropriate inclusion of cases and controls within case-control studies is essential to avoid ascertainment bias. In the pandemic context, relying on fully-controlled studies or randomly-collected samples in order to make policy decisions, e.g. analyzing the COVID-19 test results based in a random sample of people is critical to maintaining a low risk of ascertainment bias.

\begin{mydef}\textit{Availability bias}\end{mydef}
Availability bias occurs due to the natural human tendency to think that the most readily examples of things or data that come to mind are more representative than is actually the case. It can also occur in the use of artificial intelligence in healthcare if algorithms are trained or give more importance to the most readily available data which does not fully represent the target population.
Availability of information can be influenced by spin bias, biases of rhetoric, perception bias and recall bias. Confirmation bias\footnote{Confirmation bias is the most common form of bias; we all tend to see evidence in the world around us that supports our preconceived notions or hypotheses about what is going on, what is true, and what is right.} 
(when information is sought and used to support pre-existing beliefs) may lead to availability bias if data not supporting these beliefs is disregarded and not available for a particular decision or analysis.

* \textit{COVID-19 Context:} A lot of research on the impact of sex and gender on COVID-19 is based on results from most recently available data which show the similar results. However, most of these studies are mostly observational and doesn’t necessarily show a clear causal effect.  

* \textit{Preventive steps:} Availability bias is reduced or mitigated by consideration of the information and data informing any given decision and whether this is sufficient.

\begin{mydef} \textit{Chronological bias}\end{mydef}

When study participants allocated earlier to an intervention or a group are subject to different exposures or are at a different risk from participants who are recruited late.

* \textit{COVID-19 Context:} A population of men or women (being probably more exposed to COVID-19 because of work or other conditions) could have had symptoms earlier than another population and was able to be tested earlier.

* \textit{Preventive steps: }In observational data, it is very important to analyse any differences in exposure, treatment, or diagnostic criteria that may have varied over time and could affect the results.\\

\begin{mydef}
\textit{Compliance bias} 
\end{mydef}

Participants compliant with an intervention differ in some way from those not compliant which can systematically affect the outcome of interest.

* \textit{COVID-19 Context: } The adherence of a population, e.g. women to prescribed treatment, wearing mask social distancing rules more than another, e.g. men, could reduce their risk of dying from COVID-19.   

* \textit{Preventive step: } Clinical trials should attempt to collect data on compliance and while analyses should include the intention to treat analysis, where possible, exploratory and more focused secondary analyses investigating the impact of non-compliance would help inform the levels of compliance that affect the outcome of interest.

\begin{mydef}
\textit{Confounding bias}\end{mydef}
A distortion that modifies an association between an exposure and an outcome because a factor is independently associated with the exposure and the outcome.

* \textit{COVID-19 Context:} Our data shows data there is a possible correlation between smoking and death rate. Smoking can be a confounding variable. 

* \textit{Preventive step: } In observational studies associations have to be dramatic if one wants to be confident that plausible confounders have been ruled out; this is true of both beneficial and harmful associations.\\

\begin{mydef}\textit{Data-dredging bias}\end{mydef}
Data-dredging bias is a general category which includes a number of misuses of statistical inference (e.g. fishing, p-hacking) which lead to “Attractive”, but unreliable results.

* \textit{Example:} Making analytic choices (e.g. how to handle outliers, whether to combine groups, including/excluding covariates) which will produce a statistically significant p-value, making decisions about whether to collect new data on the basis of interim results; making post-hoc decisions about which statistical analyses to conduct; and generating a hypothesis to explain results which have already been obtained but presenting it as it were a hypothesis one had prior to collecting the data.

* \textit{Preventive steps: }
\begin{itemize}
\item[--] Pre-specified data analysis and collection tools.
\item[--] Better statistical education, as most of data-dredging occurs unconsciously.
\end{itemize}

* \textit{COVID-19 Context:} Most of the COVID-19 related studies are using observational data and hypothesis is formulated to explain existing results.

\begin{mydef}
\textit{Detection bias} \end{mydef}
Systematic differences between groups in accuracy of a test for a disease or how outcomes are determined. 

* \textit{Example:} Men with larger prostates are less likely to be accurately diagnosed with prostate cancer via biopsy. This minimizes the real association between obesity and prostate cancer risk.  

* \textit{Preventive steps:}
\begin{itemize}
\item[--] Identifying known factors that can affect diagnostic accuracy and statistical adjustment for detected differences.
\end{itemize}

* \textit{COVID-19 Context:} The difference in diagnostic accuracy between groups (for example, gender or race) is not known, however the possibility cannot be ruled out.  \\

\begin{mydef}
\textit{Diagnostic access bias}\end{mydef}

Individuals differ in their geographic, temporal, economic or cultural access to diagnostic procedures for a given disease.

* \textit{Example:} Health professionals are often being routinely tested, therefore healthcare workers are morel likely to be tested positive with many diseases, especially the asymptomatic cases.

* \textit{Preventive steps:}
\begin{itemize}
\item[--] Identifying known factors that can affect diagnostic access and performing sensitivity analysis to determine impact of these factors.
\end{itemize}

* \textit{COVID-19 Context:} The access to COVID-19 testing can vary drastically between the groups based on cultural, economic, professional particularities. Furthermore, the direction of bias can be inconsistent or even reversed depending on a country.  \\

\begin{mydef}
\textit{Diagnostic suspicion bias}\end{mydef}

The bias in timeliness and outcome of diagnostic tests induced by prior beliefs about the likeliness of the disease held by patients or doctors.

* \textit{Example:} Knowledge of a subject’s prior exposure to health hazards or medical personnel’s prejudices may influence both the process and the outcome of diagnostic tests.

* \textit{Preventive steps:}
\begin{itemize}
\item[--] Consecutive recruitment of patients with uniform assessment.
\item[--] Awareness of diagnostic procedures in retrospective studies.
\end{itemize}

* \textit{COVID-19 Context:} Personal or social stereotypes can affect in many ways how different groups and individuals seek or are assigned testing for COVID-19. The women in lower-income countries might not be suspected or not suspect themselves about having COVID-19, because they are staying at home. This can lead to them not being tested, even though they might have been infected, for example by male household members. 








\begin{mydef} 
\textit{Hot stuff bias}\end{mydef}

When a topic is fashionable (‘hot’) researchers and editors may be less critical and urge to publish the results.

* \textit{Example:} The bias often affect the topics that are perceived to be of imminent relevance or have a controversial aspect. 

* \textit{Preventive steps:}
\begin{itemize}
\item[--] Adherence to stringent standards for reporting and study protocols, rejecting studies whose protocols have not been published.
\item[--] Inclusion, whenever relevant, of guidelines for premature termination of studies and results release (capping the amount of support given to fashionable topics compared with important non-fashionable research)
\end{itemize}

* \textit{COVID-19 Context:} The overwhelming publicity of Corona virus related topics may affect both medical practitioners and academic community. Medical professionals may tend to pay more attention to COVID-19 patients creating disparity between them and those having other diseases. The “hot topic” effect also pressurizes researchers and editors to obtain and publish COVID-19 related study results, possibly compromising the scrutiny of the scientific process.

\begin{mydef}
\textit{Informed presence bias}
\end{mydef}

The part of population that is observed in the electronic health records is systematically different from the general population, because presence in it is affected by the person’s health status. 

* \textit{Example:} Pregnant women are exposed to medical tests much more than non-pregnant, possibly creating spurious association between pregnancy and a disease, which otherwise would remain not diagnosed. 

* \textit{Preventive steps:}
\begin{itemize}
\item[--] Awareness of limitations of healthcare records.
\item[--] Sensitivity analysis.
\end{itemize}

* \textit{COVID-19 Context:} Informed presence bias can manifest in the COVID-19 related case if data is systematically biased towards severe cases or a group which more likely to be admitted to the hospital and get tested. 

\begin{mydef}
\textit{Insensitive measure bias}
\end{mydef}

The bias arises when the tool or test outcome is not accurate.
\textit{Example:} Difference in determining stage of cancer by oncologists specializing and not specializing in the type of tumor. 

* \textit{Preventive steps:}
\begin{itemize}
\item[--] Using outcome assessment tools.
\item[--] Using validated scale measures.
\end{itemize}

* \textit{COVID-19 Context:} the tests used to diagnose COVID-19 are not uniform and differ in accuracy, furthermore some less accurate tests can be preferred in specific cases because of faster results or being less uncomfortable for the patient. This can create systematic differences among tested groups.

\begin{mydef}
\textit{Mimicry bias}
\end{mydef}

An innocent exposure may become suspicious if, rather than causing disease, it causes a benign disorder which resembles the disease.
It is important to be sure that the outcome being investigated is the true disease, and not a condition mimicking the disease, which could lead to false conclusions 

In primary care, easily missed at first conditions and undifferentiated symptoms and signs mean that immediate diagnosis is often difficult to make.

* \textit{Example}: Syphilis can mimic a wide variety of diseases in each of its phases. Another example is intussusception. 
When self-reporting is involved, as some symptoms of pneumonia also mimic viral illness, some studies had no real way to confirm such pneumonia as a symptom of a treatment (oseltamivir in this case).  

* \textit{COVID-19 Context:} Different kind of pneumonias could be associated to COVID-19 while they really are not.

* \textit{Preventive steps}:
\begin{itemize}
\item[--] Readers should be wary of studies that rely on symptoms and self-reported measures to infer critical outcomes.
\item[--] Ensure correct diagnoses of the condition are the outcome used for studies investigating causal factors. 
\item[--] Use a gold standard diagnostic reference test to confirm the diagnosis. 
\end{itemize}

\begin{mydef}
\textit{Misclassification bias}
\end{mydef}
 Occurs when a study participant is categorised into an incorrect category altering the observed association or research outcome of interest.
Non-differential misclassification occurs when the probability of individuals being misclassified is equal across all groups in the study.  Differential misclassification occurs when the probability of being misclassified differs between groups.

* \textit{COVID-19 Context}:  Missclassification errors in COVID-19 rapid tests.

* \textit{Preventive steps:}
\begin{itemize}
\item[--] To help account for mitigation bias, true disease prevalence and relation of the disease with other factors can be compared with results when disease status was determined using diagnostic codes. Two methods can be then applied:
1) Statistical techniques include differences adjustment using quantitative bias analysis (QBA) with bias parameters (code sensitivity and specificity) of varying accuracy, and 2) Disease status imputation using bootstrap methods and disease probability models. The use of one of two statistical approaches can, but does not always, reduce bias from misclassification.

\item[--] Using quantitative bias analysis will not necessarily decrease bias. QBA is dependent upon the accuracy of the data when addressing bias. 
\item[--] Use values that are measured on the population used in the study (or ones that are similar to that in the study). 
\end{itemize}

\begin{mydef}
\textit{Non-contemporaneous control bias}
\end{mydef}
Differences in the timing of selection of case and controls within a study influence exposures and outcomes, resulting in biased estimates.

* \textit{COVID-19 Context:} trying a vaccine or preventive vitamin D treatment in order to better avoid COVID-19 infection. The results of a new treatment may affect differently after a large part of the population has been already exposed, infected or vaccinated from the virus at hand.

* \textit{Preventive step:} Since the use of historical controls has shown, for instance, to produce smaller estimates of effect, and there has been cases where adjustment of outcomes for prognostic factors 
did not change the results, caution is warranted when evaluating the results of studies using non-contemporaneous controls.





\begin{mydef}
\textit{Novelty bias}
\end{mydef}
 The tendency for an intervention to appear better when it is new. It may be comprised of selection bias (e.g. participants in early trials of a medicine are more carefully selected than in later trials), positive result bias (e.g. positive results of a treatment are selectively reported when it is new and less selectively reported later) and other forms of bias such as outcome reporting bias, confirmation bias, and hot stuff bias.

* \textit{Example:} The mere appearance that a new treatment is better when it is new, may be explained by other circumstances where a new treatment actually is better when it is new (e.g., an antimicrobial treatment may actually be better when it is first introduced due to lower rates of resistance). 

* \textit{COVID-19 Context:} 
In COVID-19 vaccine development, secondary effects may be applied after a person has suffered the disease and developed some aftermath, and thus, outcomes depend on it. Another example is that the likelihood to accept a voluntary vaccine may also depend on media-based damage of the image of laboratories that got their vaccine into market first.

* \textit{Preventive steps:}
Take novelty bias into account when making decisions based on the results of new studies. Report when results are prone to suffer novelty bias. 

\begin{mydef}
\textit{Observer bias}
\end{mydef}

The process of observing and recording information which includes systematic discrepancies from the truth. 
Where subjective judgement is part of the observation, there is great potential for variability between observers, and some of these differences might be systematic and lead to bias.

* \textit{Example:} The assessment of medical images by, e.g., radiologists diagnosing COVID-19. Pointed out areas of interest in a chest X-ray may be subject to disagreement among expert radiologists \cite{tabik2020covidgr}. The same applies to diagnosis assessment. Having algorithms agreeing over what among experts clash is a challenge. Explaining to a non educated expert is also difficult, both as observer, and as expert, if the signal from an example needs to be generalized or explained.

* \textit{
Preventive steps:} Use cross-domain expert second opinions, safety checks and explainable AI methods that align the output of the model with the majority of explanations shared by experts in the domain. For the latter case, when models are used for prediction, for instance, approaches such as explainable neural-symbolic (e.g. X-NeSyL methodology and SHAP-backprop \cite{Diaz-Rodriguez21}, or human-machine teaming \cite{fompeyrine2021enhancing}) can be used.

\begin{mydef}
\textit{One-sided reference bias}\end{mydef}

When authors restrict their references to only those works that support their position, ignoring evidence not supporting their view. 

* \textit{COVID-19 Context:} One-sided citations can be used to generate information cascades resulting in unfounded authority of claims \cite{greenberg2009citation}.

* \textit{Preventive step}:
Authors of literature reviews should follow a pre-published protocol and report the searches used to find studies, the databases searched and how articles were selected for inclusion in the review.

\begin{mydef}
\textit{Outcome reporting bias}
\end{mydef}
The selective reporting of pre-specified outcomes in published clinical trials, selectively omitting or modifying outcomes of interest showed to distort the overall treatment effect. 
Adjusting for outcome reporting bias has shown to make results no longer significant. Selective reporting may cause misrepresentation of an intervention’s efficacy, skewing the public’s perception and collective scientific understanding of its benefits and harms. 

* \textit{COVID-19 Context:}  Discarding data when features are not available for an aggregated set of diverse datasets.

* \textit{Preventive steps}: 
\begin{itemize}
\item[--] Using the \textit{COMPare} Project initiative that audits outcome switching in randomised controlled trials. 
\item[--] Clinical trial investigators must maintain a policy of transparency, and inform if they change or omit planned outcomes to the reader.
\item[--] Journal editor and reviewers should compare the final studies with their protocol or registry to assess for evidence of outcome reporting bias.
\item[--] Detect it by using a clinical trial or trial registry to compare intended outcomes of interest to the analysed outcomes published in the final paper (e.g. via databases such as 
\url{https://Clinicaltrials.gov} or the WHO clinical trials Database).
\item[--] Recognize and report incomplete and biased reporting \cite{dickersin2011recognizing}.
\item[--] Instead of discarding data when features are not available for an aggregated set of diverse datasets, it can be counteracted by replacing missing features with synthetic data. 

\end{itemize}

\begin{mydef}
\textit{Partial reference bias}
\end{mydef}
A type of a verification bias (also referred to as a work-up bias, or a referral bias) that results in missing data and potential misrepresentation of the accuracy for a new test against a reference standard test.

If a diagnostic study is at high risk of partial reference bias, the results could overestimate sensitivity and specificity, and in cases, suggest a decrease in specificity of the index test.
Studies whose reference standard is expensive or invasive are prone to partial reference bias, and they introduce the potential for verification bias. 

* \textit{Example:} As many reference tests are invasive, expensive, or carry risk, 
patients 
may be less likely to pursue further tests 
if a preliminary test is negative.

* \textit{COVID-19 Context:} Chest X-ray diagnosis of different pneumonia types and COVID-19 \cite{tabik2020covidgr} are a fast, accessible and less costly alternative to PCR tests. However, they may not always benefit from accuracy alignment with gold standard tests such as PCR.

* \textit{Preventive step:} 
As obtaining a reference test in every patient may not be ethical, practical, or cost effective, 
one alternative is performing the reference test in a random sample of the study participants, or using statistical methods developed to correct for partial reference bias \cite{de2011verification}; however, this should be done with caution.

\begin{mydef}
\textit{Perception bias}
\end{mydef}

The tendency to be subjective about people and events, causing biased information to be collected in a study, or biased interpretation of a study’s results. Some subtypes are:

\begin{itemize}
\item[--] Implicit bias: individuals hold attitudes towards people, or associate stereotypes with them, without being aware of this.
\item[--] Fundamental attribution error: individuals tend to blame their failings on circumstances around them, but consider that others are responsible for their shortcomings.
\item[--] Selective perception: expectations about people or situations affect perception.
\end{itemize}

* \textit{COVID-19 Context:} 
Self-reporting of COVID-19 symptoms due to pandemics control and isolation policies.

* \textit{Preventive step:} Interpret with caution studies which rely on self-reported information or non objective measures, which are less likely to suffer from perception bias.

\begin{mydef}
\textit{Performance bias}
\end{mydef}

Systematic differences in the care provided to members of different study groups other than the intervention under investigation.

Due to knowledge of interventions allocation, in either the researcher or the participant, this bias may inflate the estimated effect of the intervention, particularly in trials with subjective outcomes.  It occurs in trials where it is not possible to blind participants and/or researchers, such as trials of surgical interventions, nutrition or exercise. 

* \textit{Example:} The lack of double blinding (on both participant and researchers) compared to studies with clear double-blinding showed effect estimates on average 13\% higher \cite{savovic2012influence}.

* \textit{COVID-19 Context:} Self-knowledge of being diagnosed as \textit{positive} may increase subjective self-reporting of symptoms (e.g. pain).

* \textit{Preventive steps}: 
\begin{enumerate}
\item[--] If participant blinding is not feasible, the effects of this bias can be mitigated using objective outcomes. 
\item[--] If subjective outcomes are used in a trial, performance bias can be mitigated by blinding the outcome evaluator.
\end{enumerate}

\begin{mydef}
\textit{Popularity bias}
\end{mydef}
 Differences in the uptake of healthcare as a result of a public interest in a disease or condition and its possible causes results in a biased study sample.
This bias can arise from increased awareness of a certain disease or condition by the general population, or by media articles or health services as instigators of increased interest/popularity in the condition and its possible causes. 

* \textit{Example:} A tendency for more popular items to be recommended more frequently than less popular ones. 

* \textit{COVID-19 Context:} An authority going public about certain condition may shoot such popularity, e.g. politicians getting access to a particular drug.

* \textit{Preventive steps}: 
\begin{enumerate}
\item[--] Perform randomization with proper allocation concealment to prevent popularity bias.
\item[--] Report for potential selection bias and how it has been controlled. 
\item[--] Analyze data over longer terms to smooth out any variability in uptake.
\end{enumerate}
 
\begin{mydef}
\textit{Positive results bias}\end{mydef}
 The tendency to submit, accept and publish positive results rather than non-significant or negative results.

* \textit{Example:} The culture of non-publication of negative results leads to research wastage, as researchers may unnecessarily repeat studies because the results are unpublished. Outcome reporting bias exists, impacting on the pooled summary in systematic reviews, which are published faster compared with trials with negative outcomes \cite{song2010dissemination}. 

* \textit{COVID-19 Context:} The needs for pandemics and disaster response blooms research. Therefore the rush to provide solutions may compromise the quality of critical steps of the research and analysis process.

* \textit{Preventive steps}:   
\begin{enumerate}
\item[--] Systematic reviews should attempt to find and include all relevant studies that are difficult to find and should report statistical methods to deal with the presence of unpublished results.
\item[--] Awareness of publication bias amongst the research and publishing communities should help reduce the issue of negative results being rejected by journals. 
\end{enumerate}

\begin{mydef}
\textit{Prevalence-incidence (Neyman) bias}
\end{mydef}

Exclusion of individuals with severe or mild disease resulting in a systematic error in the estimated association or effect of an exposure on an outcome.

On one hand, excluding patients who have died will make the disease appear less severe. On the other hand, excluding patients who have recovered will make the disease seem more severe. The longer the time between exposure and investigation, the larger the likelihood of individuals dying or recovering from the disease being excluded from the analysis. Thus, this bias more largely impacts long-lasting diseases than short ones.

* \textit{COVID-19 Context:} Studies on the influence of COVID-19 effects indicate deficit of vitamin D as a factor. Can we measure the effects on supplements on patients of oldest ages when they are the population most affected in terms of deaths?

* \textit{Preventive step:} Assess outcomes on a broader age groups covering a longer and broader population range.


\begin{mydef}
\textit{Previous opinion bias}
\end{mydef}

The results of a previous assessment, test result or diagnosis, if known, may affect the results of subsequent processes on the same patient.

* \textit{COVID-19 Context:} Because of the randomness in the symptoms of COVID-19, physicians are not able to understand the patterns of COVID-19. 
The failure of fast tests such as serologies may avoid necessary PCR tests, as well as create false negatives.
Likewise, in early self-detection, a patient might change to a different health state but due to the similarity of symptoms with other conditions, a patient can be declared as a COVID-19 positive, resulting in false positives.

* \textit{Preventive steps:}
\begin{itemize}
\item[--] Every researcher and healthcare practitioner must strive to observe and use the best available date, in the best possible way, while being aware that one’s preconceptions can be misleading.
\item[--] Healthcare professionals should often consult colleagues for a second opinion and ask open-ended questions. Performing questions such as \textit{"Could you examine this patient? – I need a second opinion"} can minimize previous opinion bias. 
\end{itemize}







\begin{mydef}
\textit{Referral filter bias}\end{mydef}
Referral of any group of unwell people from primary to secondary to tertiary care, causing an increase in the concentration of rare cases, more complex cases or people with worse outcomes.

* \textit{COVID-19 Context:} COVID-19 patients may have different characteristics compared to the studies in the hospital. If patients are more likely to have unusual features -to have more severe high blood pressure, cardiovascular disease, different lifestyles (smoking, drinking), and have secondary causes-, they may not, therefore, represent the population with COVID-19 as a whole.

* \textit{Potential solution:} The minimizing selection bias can also take into account referral filter bias. And the interpretation of study results should always consider whether the inclusion criteria or sampling frame, enable generalizable to the target population.

\begin{mydef} 
\textit{Reporting bias}\end{mydef}
A systematic distortion that arises from the selective disclosure or withholding of information by parties involved in the design, conduct, analysis, or dissemination of a study or research finding.

* \textit{COVID-19 Context:} A vulnerability group to COVID-19 may be found in a study, and its reporting may be delayed/ hidden to avoid healthcare system collapse and other economical losses.

* \textit{Preventive steps:} Transparency is necessary for medical research. Moreover, tools have been developed to check for reporting bias including the Cochrane Risk of Bias Tool, GRADE and ORBIT-II. 

\begin{mydef} 
\textit{Selection bias} \end{mydef}
It occurs when individuals or groups in a study differ systematically from the population of interest leading to a systematic error in an association or outcome. 

* \textit{COVID-19 Context:} Several studies show that males are more vulnerable to COVID-19. Moreover, aged people have more risk of COVID-19. However, more research requires to be on how sex intersects with age and increases the risk of severe COVID-19.

* \textit{Preventive steps:} To assess the probable degree of selection bias, authors should include the following information at different stages of the trial or study:
\begin{itemize}
\item[--] Number of participants screened as well as randomised/included.
\item[--] How intervention/exposure groups compared at baselines.
\item[--] To what extent potential participants were re-screened.
\item[--] Exactly what procedures were put in place to prevent prediction of future allocations and knowledge of previous allocations.
\item[--] What the restrictions were on randomisation, e.g. block sizes.
\item[--] Any evidence of unblinding\footnote{i.e., re-identification of the treatment code of a subject in studies where the treatment assignment was unknown to the subject and investigators.}.
\item[--] How missing data from participants lost to follow-up were handled.
\end{itemize}

\begin{mydef}
\textit{Spectrum bias}
\end{mydef}
Occurs when a diagnostic test is studied in a different range of individuals to the intended population for the test. 

It is desirable that diagnostic tests have both high sensitivity (the proportion of people testing positive who actually have the disease) and high specificity (the proportion of people testing negative who do not have the disease). Spectrum bias is one of the factors affecting sensitivity and specificity of a test when applied on a different mix of patients. For instance, there is evidence that when using a case-control design in diagnostic accuracy studies, both sensitivity and specificity are increased.

* \textit{Example:} facial recognition systems are less accurate over black people, because Caucasian people are over-represented.

* \textit{COVID-19 Context:} In the world, different testing strategies are using to check the COVID-19 that are not accurately proved. As the testing strategies are varying the COVID-19 reports are also different from the other testing strategies. 


* \textit{
Preventive steps:} Systematic reviews of diagnostic accuracy should include pre-specified subgroup analyses of all clinically relevant patient characteristics. Furthermore, when using a particular diagnostic test, clinicians should be aware of the differences in characteristics between the patient or population in front of them that must include in the research.

* \textit{Preventive steps:} 
\begin{itemize}
    \item[--] Studies of diagnostic test accuracy should recruit individuals that are representative of the population in which the test will be typically used. 
    \item[--] Follow STARD guidelines and PRISMA for diagnostic test accuracy. 
    \item[--] Compute the predictive capacity of the classifier, learn invariant feature representations, apply empirical risk minimization (ERM)\footnote{(ERM is a classic idea from survey theory that tries to match the training to the test data. It is challenging to demonstrate, since data is weighted.} \cite{donini2018empirical} to biased data. Survey schemes can also accelerate the learning by using appropriate batch sizes. Importance sampling and censorship probabilities are strategies to deal with biased training samples available. Because facial recognition algorithms cannot be perfectly fair
    \cite{maxwell2020facial}, fairness constraints must be incorporated to the ERM program. 
    \item[--] Any global testing strategy must be applied in all over the world which is accepted by the WHO and which gives more accurate a bias-free and fast results.
\end{itemize}

\begin{mydef}
\textit{Spin bias}\end{mydef}

The intentional or unintentional distorted interpretation of research results, unjustifiably suggesting favourable or unfavourable findings that can result in misleading conclusions. 
Spin in clinical research reports consists of presenting a study in a more positive way than the actual results reflect or down playing harms \cite{mahtani2016spin}. Examples of errors include: Implying significance by referring to trends, Inferring significance from statistical differences in subgroups, or stressing per protocol rather than intention to treat analysis, among many others.
Spinning can harm news and media interest \cite{adams2019claims}. 

* \textit{Preventive steps:} 
\begin{itemize}
    \item[--] Peer reviewing and journal editorial checking for result significance and causal signals rather than correlational results should be done.
    \item[--] Journals should include full data sets (where ethically appropriate), to promote reanalysis and replication.
    \item[--] Use validated spin detection tools. 
\end{itemize}

* \textit{Example:} misleading reporting reduces completeness, transparency and the value of reports of health research.

* \textit{COVID-19 Context:} Different COVID-19 patients may belong to different subgroups with various symptoms and diseases that consequently result in nonsignificant reports that can mislead the reader to perform/take experimental decisions.

* \textit{Preventive steps:} 

\begin{itemize}
    \item[--] Follow the EQUATOR network\footnote{\url{http://www.equator-network.org/wp-content/uploads/2011/10/Research-publication-spin.pdf}} practices to avoid the (non-exhaustive list of) errors above.
    \item[--] Evidence used for decision making should not be based on a single study, but instead, be informed by systematic reviews of relevant research.
\end{itemize}

\begin{mydef}
\textit{Starting time bias}\end{mydef}

Arises when there is a failure to identify a common starting time for an exposure or a disease.
Failing to identify a common starting time for exposure or illness may lead to systematic misclassification (relates to selection bias,  non-contemporaneous control bias and could be an example of chronological bias).

* \textit{COVID-19 Context:} 
\begin{itemize}
\item[--] COVID-19 PCR negative tests as a consequence of being done too early. 
\item[--] 
During the starting phase of COVID-19 a person was kept in observation for 7 days after being a contact case, to make sure whether that person is infected of COVID-19. As COVID-19 research expanded, the observation days were also increased depending on the patient situation. Results and contact tracing protocols thus need to take this variable into account.
\end{itemize}

* \textit{
Preventive steps}: Cross-border contact tracing apps 
should be used to estimate indoor wifi positioning, outdoor GPS-based, and bluetooth level accuracies. Since the aerosols responsible for human spread last for hours in the air, reverse engineering of positive cases should allow for an average infection time, place, and disease detectability for optimal test taking. 

\begin{mydef}
\textit{Substitution game bias}\end{mydef}

Occurs when doing a substitution of the clinically important endpoint, or an exposure, with a surrogate marker for the disease. 
Surrogate outcomes in trials are used as both substitution and for predicting the effect of a treatment in the absence of data on patient relevant outcomes. 

* \textit{Example}: On average, trials using surrogate primary outcomes report larger treatment effects than matched ones using primary outcomes \cite{ciani2013comparison}. 
Thus, surrogate markers should be interpreted with additional caution.

* \textit{COVID-19 Context:}
It may be applied to less often COVID-19 symptoms such as the type of nasal congestion, area of eczemas or the amount of time smell and taste are lost. 

* \textit{Preventive steps:} the effect of the exposure or intervention on the surrogate marker should predict the effect on the clinical outcome (not merely be correlated).

\begin{mydef}
\textit{Unacceptability bias}\end{mydef}

A systematic difference in response rates or uptake of tests due to their “unacceptability”.
Due to perception of too invasive questions, factors such as the use of illegal drugs, sexual behaviour or other vices can be associated with under-reporting. For instance, alcohol consumption is underestimated. 

* \textit{Example}: Under-reporting can be a result of stigma of a diagnosis or illness, e.g., in relation to work prospects. 

* \textit{COVID-19 Context}: 
\begin{itemize}
\item[--] As we know, data is not collected efficiently with proper planning and administration. This may not lead to being appropriate to study COVID-19 patients.
\item[--] A person who reported positive with some common symptoms on the first day may have different symptoms after five days with some new diseases due to lack of appropriate study and data.
\end{itemize}

* \textit{Preventive steps:} 
\begin{itemize}
\item[--] Make questionnaires anonymous and place unacceptable questions toward the end so they do not affect other questions. 
\item[--] Avoidance of harm should be a priority in any research work. 
\item[--] The collection of sensitive or personal information has ethical implications and thus its administration must be planned accordingly, and be appropriate to the needs of study participants.
\end{itemize}

\begin{mydef}
\textit{Unacceptable disease bias}\end{mydef}

Lower rates of reporting of certain \textit{unacceptable} diseases compared with other health conditions. 
This bias is similar to unacceptable exposure bias, as individuals may under-report exposures regarded as socially undesirable. Unacceptable disease bias, a subtype of reporting bias, due to shame or embarrassment, results in data not entirely reported.  This reduces the validity of data on such conditions and may make the comparison in studies difficult.

* \textit{Example:} Examples reducing the validity of data on such conditions, making the comparison in studies difficult are the reputation of, e.g. an elderly home/residence may have or economic development dependent services such as hostelry (hospitality, restauration services and more).

* \textit{COVID-19 Context}: Under-reporting can be a result of stigma of a diagnosis or illness, e.g., in relation to work prospects. 

* \textit{Preventive steps}: Make questionnaires anonymised and place unacceptable questions toward the end so they do not affect other questions. 

\begin{mydef}
\textit{Unmasking (detection signal) Bias}\end{mydef}

An innocent exposure that, rather than causing a disease, causes a sign or symptom that precipitates a search for the disease.

* \textit{Example}:  If a medication can cause vaginal bleeding, and people with this symptom go sooner to the doctor and receive earlier or more intensive examination and investigations to diagnose cancer, it may appear that the medication caused the cancer. However, all that may have happened is that the medication has prompted an earlier or more intensive search for the disease, leading to an apparently increased rate among those using the medication.

* \textit{Example:} An earlier screening policy for COVID-19 testing (e.g. among healthcare or teaching staff) that is performed more massively or efficiently may uncover a larger amount of infected population, and thus, implement a contention policy (lockdown) earlier, stopping in this moment the additional tests unless symptoms are shown. Such policies may ease the under-reporting of a large amount of asymptomatic population. 

* \textit{COVID-19 Context:} pregnant women or chronic patients that often are checked-up may be tested more often to avoid visits to the hospital when infected by a virus. This may count them as more prone to have the disease, while other asymptomatic patients may go overlooked.

* \textit{Preventive steps}:
\begin{itemize}
    \item[--] Normalize data by balancing the periods accounted for, to include an equal amount of time intervals for all groups. 
    \item[--] Account intervals of data collection to begin at a time with a given condition. 
\end{itemize}

\begin{mydef}
\textit{Verification bias}\end{mydef}

It occurs when only a proportion of the study group receives confirmation of the diagnosis by the reference standard, or if some patients receive a different reference standard at the time of diagnosis. Prone to verification bias are studies where the reference standard was an expensive or invasive test. 

* \textit{COVID-19 Context}: Data malleability or contradicting data: Diagnose tests change results as a follow up, and past test may result on wrong information due to wrong timing in testing.

* \textit{Preventive steps}:
\begin{itemize}
    \item[--] Allow diagnosis updates in the data post-analysis. 
    \item[--] In order to not take cheaper tests for granted, if their accuracy is not as high as the most expensive ones (e.g. serology vs PCR test), cheaper screening methods are sought that do not compromise accuracy.
    \item[--] Since tests may be taken at non optimal times (during incubation of the disease, e.g., still providing a negative PCR test), even the most accurate tests should provide an uncertainty window. 
    \item[--] As COVID-19 test systems are different all over the world, the reliability of their results also differs. Therefore, the world has different ways of verifying and certifying the COVID-19.
    \item[--] Participant non availability in a follow up study: Older participants or internationally traveller citizens that get tested at different countries with different diagnosis protocols may have different outcomes and be reflected differently in the counting. Furthermore, due to each country having a different COVID-19 tracker mobile app. that acts as an information silo (sharing no data across countries), these are no longer tracked when borders are crossed. Patients dying may also impede the study to conclude with the associated results.
\end{itemize}

* \textit{Preventive steps: } 
\begin{itemize}
    \item[--] Allow for an international citizen ID that tracks movement, diagnosis and diagnosis incompatibilities cross-borders.
    \item[--] Finding counterfactual or contrastive explanations for a model's functioning based on patients that died too early to continue within the study, on patients that share physiological features similarly enough for some kind of uncertainty-informed estimated effects.
\end{itemize}

\begin{mydef}
\textit{Volunteer bias}\end{mydef}

Participants volunteering to take part in a study intrinsically have different characteristics from the general population of interest.

* \textit{COVID-19 Context:} Only participants unemployed or with perfect mobility patterns (i.e., young) and with economic difficulties (i.e., without access to healthcare insurance) volunteer to be tested on a new vaccine or treatment which may have unknown secondary effects. Their health situation may also be different to other people that would not have time to be part of that research. 

* \textit{Preventive step:} For the dataset to be representative, samples must be complemented on participants from the missing (e.g. age, health status) range.


\subsection{The MoSCoW scale for task prioritizing}

The MoSCoW Scale\footnote{\url{https://www.kecg.co/blog/2018/7/22/task-prioritisation-hack-using-moscow-method}} 
is a popular prioritization technique used in software engineering for managing requirements that stands for 4 different categories of initiatives: \textit{must-haves, should-haves, could-haves}, and \textit{will not have at this time}. It prioritises any task and study, and can be applied to requirements, products, use cases, user stories, acceptance criteria and tests.

\bibliographystyle{abbrv} %
\bibliography{refs}

\end{document}